\documentclass[aps,prb,reprint,superscriptaddress]{revtex4-2}

\usepackage{amsmath, amssymb}
\usepackage{bm}
\usepackage{graphicx}
\usepackage{color}
\usepackage{dcolumn}
\usepackage{bm}
\usepackage[hidelinks]{hyperref}
\hypersetup{
  colorlinks   = true, 
  urlcolor     = blue, 
  linkcolor    = blue, 
  citecolor   = red 
}
\usepackage{multirow}

\allowdisplaybreaks[1]

\begin{document}

\title{
Vibronic order and emergent magnetism in cubic $d^1$ double perovskites
}

\author{Naoya Iwahara}
\email[]{naoya.iwahara@gmail.com}
\affiliation{Graduate School of Engineering, Chiba University, 1-33 Yayoi-cho, Inage-ku, Chiba-shi, Chiba 263-8522, Japan} 

\author{Liviu F. Chibotaru}
\email[]{liviu.chibotaru@kuleuven.be}
\affiliation{Theory of Nanomaterials Group, KU Leuven, Celestijnenlaan 200F, B-3001 Leuven, Belgium}

\begin{abstract}
The synergistic interplay of different interactions in materials leads to the emergence of novel quantum phenomena.  
Spin-orbit and vibronic couplings usually counteract each other, however, in cubic $d^1$ double perovskites they coexist and give rise to spin-orbit-lattice entanglement with unquenched dynamic Jahn-Teller effect on the metal sites.
The correlation of these entangled states induced by intersite interactions was not assessed so far. 
Here, we investigate the joint cooperative effect of spin-orbit and vibronic interactions on the formation of the ordered phases in $d^1$ double perovskites.
We found that the magnetically ordered states in these systems coexist with a dynamic vibronic order characterized by the ordering of vibronic quadrupole moments on sites. 
This treatment allows the rationalization of a number of unexplained features of experimentally investigated phases.
\end{abstract}

\maketitle

\begin{figure}[bt]
\centering
\begin{tabular}{lllll}
(a) &~& (b) &~& (c)\\
 \includegraphics[bb = 0 0 552 522, height=0.27\linewidth]{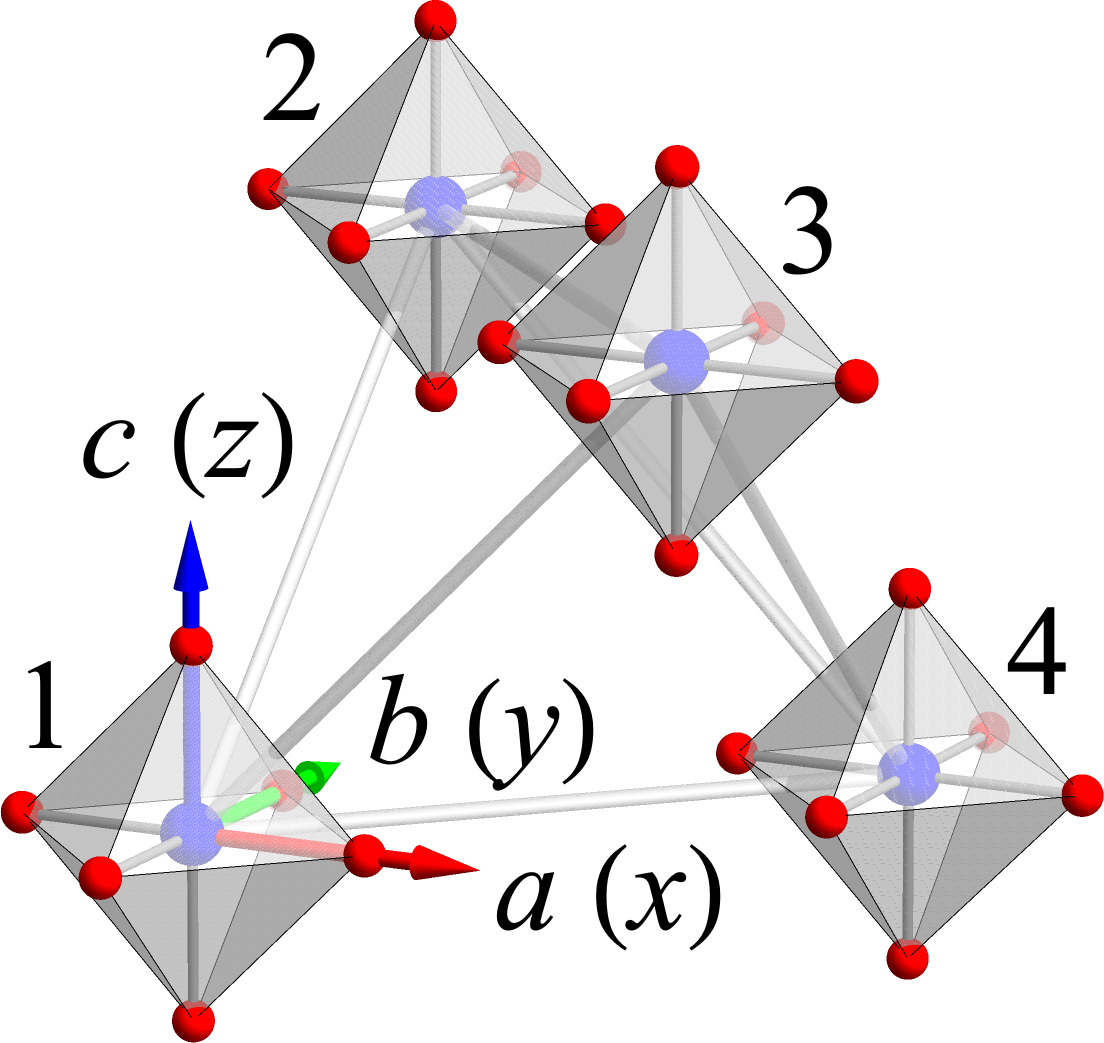}
& & 
\includegraphics[height=0.34\linewidth]{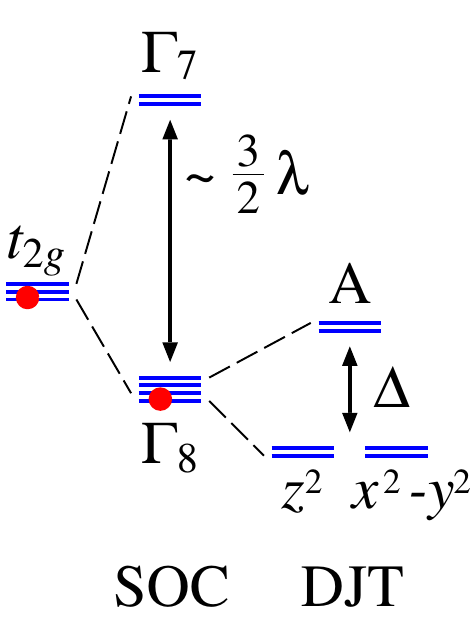}
& &
\includegraphics[bb = 0 0 877 872, height=0.32\linewidth]{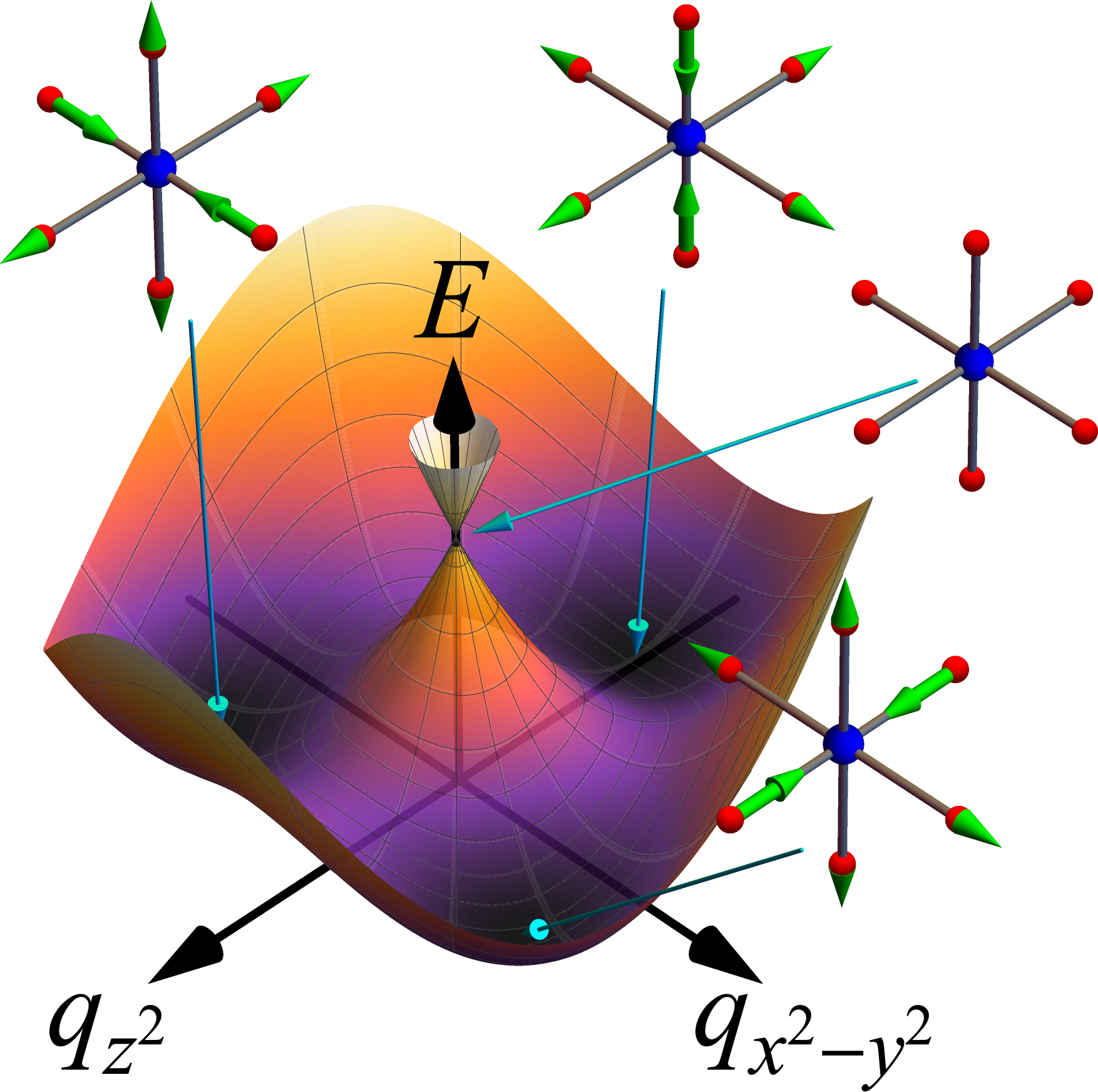}
\end{tabular}
\caption{
Structure of $d^1$ double perovskites and local quantum states. 
(a) Conventional cell of fcc lattice of octahedral centers. 
Blue and red spheres are metal and ligand atoms.  
(b) The energy diagram of $d^1$ site. 
(c) APES with respect to JT-active normal coordinates.
}
\label{Fig:JT}
\end{figure}

{\it Introduction.---}
In Mott insulators, spin-orbit entanglement is a source of nontrivial magnetism. 
The anisotropic exchange interaction induced by strong spin-orbit coupling in heavy transition metal compounds has been intensively employed to realize Kitaev spin liquid phase \cite{Rau2016, Takagi2019}. 
When the total angular momentum on metal sites amounts to $J_\text{eff} > 1/2$, the magnetic interaction is not only anisotropic but also multipolar.
A well-known example of magnetic multipolar systems is a family of geometrically frustrated $4d^1/5d^1$ double perovskites with $J_\text{eff} = 3/2$ [Fig. \ref{Fig:JT}] characterized by unusual order 
\cite{Stitzer2002, Yamaura2006, Erickson2007, Steele2011, Marjerrison2016, Lu2017, Liu2018, Willa2019, Ishikawa2019, Hirai2019, Hirai2020, Hirai2021, Ishikawa2021, Ishikawa2021b, Arima2022, Woodward2022, Cong2022, Tehrani2023} 
and glassy phases \cite{Cussen2006, Aharen2010, deVries2010, Coomer2013, Lee2021, Mustonen2022}.

The origin of the ordered phases in these systems is still puzzling. 
These compounds exhibit either canted ferromagnetic (FM110) or antiferromagnetic (AFM) phases. 
Spin-orbit based theories \cite{Chen2010, Romhanyi2017, Svoboda2021} predict the main features of the FM110 phase, while showing discrepancies with experiment, particularly, in the following two cases \cite{Ishikawa2019, Hirai2020}. 
(1) A high-resolution x-ray scattering study on single-crystalline Ba$_2$MgReO$_6$ revealed the coexistence of antiferro $x^2-y^2$ (rhombic deformations on sites) and ferro $z^2$ (tetragonal elongations on sites)
quadrupole orders in the high-temperature phase \cite{Hirai2020}. 
(2) A family of tantalum compounds, $A_2$Ta$X_6$ ($A =$ Rb, Cs, $X = $ Cl, Br), exhibits AFM order that accompanies ferro $z^2$ quadrupole order of tetragonal compression type \cite{Ishikawa2019, Ishikawa2021}. 
However, conventional spin-orbit theories do not predict quadrupole orders which would match the observed structural distortions 
\cite{Chen2010, Romhanyi2017, Svoboda2021}.

In the $d^1$ double perovskites, the vibronic coupling at each metal site gives rise to dynamic Jahn-Teller (JT) effect resulting in spin-orbit-lattice entanglement on sites \cite{Iwahara2018}. 
Indeed, the {\it ab initio} calculations for molybdenum and osmium double perovskites proved that the $J_\text{eff}=3/2$ states are strongly coupled to the JT-active modes \cite{Iwahara2018, Xu2016}.
The resulting dynamic JT stabilization is much larger than the magnetic interaction quantified by the Curie-Weiss constants, indicating that the dynamic JT effect persists in these crystals given the lack of common atoms for different metal octahedra [Fig. \ref{Fig:JT}(a)]. 
The dynamic JT effect smears out the structural anisotropy, explaining why many $d^1$ double perovskites are cubic \cite{Cussen2006, Erickson2007, Aharen2010, Coomer2013, Woodward2022, Cong2022, Mustonen2022}, while under external field the ground states show slight localization at a JT deformed geometry \cite{Iwahara2018}.
Despite the importance of the dynamical JT effect in $d^1$ double perovskites, it was still not properly treated for the description of the magnetic phases in these compounds. 

In this work, we extend the vibronic treatment of individual metal sites over the cooperative effect of joint spin-orbit and dynamic Jahn-Teller interactions in a family of cubic $d^1$ double perovskites. 
We found that elastic and exchange couplings between entangled spin-orbit vibronic states on metal sites give rise to rich phases of coexisting magnetic and vibronic orders.

{\it Microscopic model for the $d^1$ compounds.---}
In $d^1$ double perovskites, the metal octahedra form a face-centered-cubic (fcc) lattice [Fig. \ref{Fig:JT}(a)].
In each $d^1$ octahedron, the $d$ orbitals split into the $e_g$ doublet and $t_{2g}$ triplet, and an electron occupies the $t_{2g}$ orbitals in the low-energy states. 
The microscopic model for the $t_{2g}$ electrons comprises intrasite bielectronic, spin-orbit and Jahn-Teller, and intersite electron transfer and elastic interactions:
\begin{align}
 \hat{H} &= \sum_i \left(\hat{H}_U^i + \hat{H}_\text{SO}^i + \hat{H}_\text{JT}^i \right)
 + \sum_{i<j} \left(\hat{H}^{ij}_t + \hat{H}^{ij}_\text{vib} \right).
 \label{Eq:H}
\end{align}

The Coulomb and electron transfer interactions are 
\begin{align}
 \hat{H}_U^i &= 
   \sum_{\gamma} U \hat{n}_{i\gamma\uparrow} \hat{n}_{i\gamma\downarrow} 
 + \sum_{\gamma < \gamma'} \sum_{\sigma\sigma'} 
 \Bigg[
 (U-2J_H) \hat{n}_{i\gamma\sigma} \hat{n}_{i\gamma'\sigma'}
 \nonumber\\
 &+
  J_H
 \hat{a}_{i\gamma\sigma}^\dagger
 \hat{a}_{i\gamma'\sigma'}^\dagger
 \hat{a}_{i\gamma\sigma'}
 \hat{a}_{i\gamma'\sigma}
 \Bigg]
 \nonumber\\
 &+ 
 \sum_{\gamma \ne \gamma'} J_H 
 \hat{a}_{i\gamma\uparrow}^\dagger 
 \hat{a}_{i\gamma\downarrow}^\dagger 
 \hat{a}_{i\gamma'\downarrow}
 \hat{a}_{i\gamma\uparrow},
 \label{Eq:HU}
\\ 
\hat{H}_t^{ij} &= \sum_{\gamma\gamma'} \sum_\sigma t^{ij}_{\gamma\gamma'} 
\left(\hat{a}_{i\gamma\sigma}^\dagger \hat{a}_{j\gamma'\sigma}
+
\hat{a}_{j\gamma'\sigma}^\dagger \hat{a}_{i\gamma\sigma} \right).
\label{Eq:Ht}
\end{align}
where $\gamma$ ($= yz, zx, xy$) are three $t_{2g}$ orbitals at each site $i$, 
$U$ and $J_H$ are the Coulomb and Hund coupling parameters, respectively, and $t^{ij}_{\gamma\gamma'}$ is the electron transfer parameter between orbitals $\gamma$ and $\gamma'$.
We assume that $U/t$ is sufficiently large for the development of the Mott insulating phase.

The spin-orbit coupling at each metal site can be expressed via
the effective orbital angular momentum $\tilde{l}=1$ of the $t_{2g}$ shell [\S 7.1.1 in Ref. \cite{Sugano1970}] as follows,
\begin{align}
 \hat{H}_\text{SO}^i &= 
 \lambda \tilde{\bm{l}}^i \cdot \hat{\bm{s}}^i.
 \label{Eq:HSO}
\end{align}
It stabilizes $J_\text{eff} = 3/2$ ($\Gamma_8$) at each site given $\lambda>0$ [Fig. \ref{Fig:JT}(b), \S 7.1.2 in Ref. \cite{Sugano1970}].

The $t_{2g}$ orbitals also interact with the JT active $E_g$ vibrations of the octahedron [\S 3.3.2 in Ref. \cite{Bersuker1989}]:
\begin{align}
 \hat{H}_\text{JT}^i &= \sum_\alpha \frac{\hslash \omega}{2} (p_{i\alpha}^2 + q_{i\alpha}^2) 
\nonumber\\
&+
\hslash \omega g 
\left[
\left( q_{iz^2} + \eta \{q^2_i\}_{z^2} \right) \hat{P}_{xy}^i + \text{cycl.} \right],
\label{Eq:VJT1}
\end{align}
where $q_{\alpha}$ and $p_\alpha$ ($\alpha = z^2, x^2-y^2$) are dimensionless normal coordinate for the JT active modes and conjugate momentum, respectively,
$\{q^2\}_{z^2}$ is the symmetrized quadratic polynomial of $q$, 
$\omega$ is the frequency of the JT active mode, 
$g>0$ and $g\eta>0$ are the dimensionless linear and quadratic vibronic coupling parameters, respectively,
$\hat{P}^i_{\gamma}$ is the projection operator into the $\gamma$ orbital on site $i$, and 
the ``cycl.'' indicates cyclic permutations of $x,y,z$.
$q_{z^2} > 0$ ($< 0$) corresponds to tetragonal elongation (compression) of the octaheron [Fig. \ref{Fig:JT}(c)].
The adiabatic potential energy surface (APES) from Eq. (\ref{Eq:VJT1}) consists of three independent paraboloids [Fig. 3.6 in Ref. \cite{Bersuker1989}].
Each paraboloid corresponds to one $t_{2g}$ orbital and has a minimum with tetragonal compression normal to the plane of this orbital.

Finally, the JT active vibrations on neighbor sites interact through the elastic coupling:
\begin{align}
\hat{H}_\text{vib}^{ij} &= (q_{iz^2}, q_{ix^2-y^2}) \bm{D}_0^{ij}(\theta) (q_{jz^2}, q_{jx^2-y^2})^T.
\label{Eq:Hvib}
\end{align}

The strongest is the on-site Coulomb interaction ($U \approx$ 3 eV \cite{Erickson2007, Mosca2023}), followed by spin-orbit coupling ($\lambda \approx$ 0.25-0.35 eV for $5d^1$ ions \cite{Iwahara2018, Ishikawa2019}) 
and then Jahn-Teller ($\hslash \omega g \approx$ 50 meV \cite{Iwahara2018}) and electron transfer ($t\approx$ 50 meV \cite{Erickson2007}) interactions.
The intersite elastic interaction of JT active vibrations is about one order of magnitude smaller than their frequency, $\hslash \omega \approx$ 50 meV.

We treat the intrasite interactions exactly and the intersite interactions perturbatively following the standard way for correlated insulators. 
We reduce $\hat{H}_U$ and $\hat{H}_t$ to the spin-orbital superexchange interaction for the description of the low-energy phenomena within the Mott insulating phase. 
The magnitude of the exchange interaction parameter, $J \approx t^2/U \approx$ 1 meV, is consistent with the experimental estimates from the Curie-Weiss constants $J \approx \Theta/z$, where $z=12$ is the number of the nearest-neighbor sites.
The exchange interaction is by two orders of magnitude smaller than the intrasite interactions and is comparable to the intersite elastic interaction. 
Below, we describe the interactions in descending order of the energy scale. 

{\it Local spin-orbital-lattice entangled states.---}
The octahedral symmetry of each $d^1$ site enables the persistence of the JT effect in strong spin-orbit coupled states. 
The spin-orbit $\Gamma_8$ quartet splits into two Kramers doublets by JT deformations [Fig. \ref{Fig:JT}(c), \S 3.3.3 in Ref. \cite{Bersuker1989}]:
Under tetragonal compression along an axis $\gamma$ ($= x,y,z$), the lower energy Kramers pair contains a dominant contribution from the $t_{2g}$ orbital lying in the plane perpendicular to the $\gamma$ axis.

The nature of the JT coupling within the $\Gamma_8$ quartet becomes transparent by introducing pseudo orbital and pseudo spin. 
Because of the relation $\Gamma_8 = \Gamma_3 \otimes \Gamma_6$, we can represent the $\Gamma_8$ multiplet via a direct product of $e$ ($\Gamma_3$) `pseudo orbital' $\tilde{\bm{\tau}}$ and $\Gamma_6$ `pseudo spin' $\tilde{\bm{s}}$ ($\tilde{\tau} = \tilde{s} = 1/2$) \cite{Natori2016, Romhanyi2017}: 
$|\Gamma_8,\mp \frac{1}{2}\rangle = \pm |\tilde{\tau}_z = +\frac{1}{2}, \tilde{s}_z = \mp \frac{1}{2} \rangle$ and $|\Gamma_8, \mp \frac{3}{2}\rangle = \pm|\tilde{\tau}_z = -\frac{1}{2}, \tilde{s}_z = \pm \frac{1}{2}\rangle$.
$\tilde{\tau}_{z(x)}$ is an electric $z^2$ ($x^2-y^2$) quadrupole moment operator. 
Within this representation, the JT coupling term in Eq. (\ref{Eq:VJT1}) reads as  
\begin{align}
- \hslash \omega g 
 \left[
 \left( {q}_{z^2} + \eta \{{q}^2\}_{z^2} \right) \tilde{\tau}_z
  +
 \left( {q}_{x^2-y^2} + \eta \{{q}^2\}_{x^2-y^2} \right) \tilde{\tau}_x
 \right].
 \label{Eq:VJT2}
\end{align}
The pseudo-JT coupling between spin-orbit multiplets and the anharmonicity effects vary the magnitude of the warping of the APES [Fig. \ref{Fig:JT} (b), (c)].

The local quantum states are of vibronic type characterized by spin-orbit and lattice entanglement \cite{Iwahara2018}. 
We start with three localized states around the minima of the APES with $\gamma$ ($=x,y,z$) compression, $|\Phi_\gamma\rangle$ [Fig. \ref{Fig:JT}(c)].
A localized state $|\Phi_\gamma\rangle$ is the direct product of the pseudo orbital state $\cos \phi_\gamma |\tilde{\tau}_z = -\frac{1}{2}\rangle + \sin \phi_\gamma |\tilde{\tau}_z = +\frac{1}{2}\rangle$ and the ground vibrational state at the minimum, where $\phi_\gamma = -\frac{\pi}{6}, \frac{\pi}{6}, \frac{\pi}{2}$ for $\gamma = x,y,z$, respectively. 
The kinetic energy term in Eq. (\ref{Eq:VJT1}) promotes the delocalization over other minima [\S 4.3.3 in Ref. \cite{Bersuker1989}]. 
Then the vibronic states are linear combinations of $|\Phi_\gamma\rangle$'s, 
two ground states are $E$ type, 
\begin{align}
|z^2\rangle &= \frac{1}{\sqrt{6}}(2|\Phi_z\rangle - |\Phi_x\rangle - |\Phi_y\rangle),
\nonumber\\
|x^2-y^2\rangle &= \frac{1}{\sqrt{2}}(|\Phi_x\rangle - |\Phi_y\rangle),
\label{Eq:E}
\end{align}
and one excited state of $A$ type, 
\begin{align}
|A\rangle = \frac{1}{\sqrt{3}}(|\Phi_x\rangle + |\Phi_y\rangle + |\Phi_z\rangle),
\label{Eq:A}
\end{align}
defining a one-site Hamiltonian, 
\begin{align}
\hat{\mathcal{H}}_0 = \Delta \hat{\mathcal{P}}_{A},
\label{Eq:H0}
\end{align}
where $\hat{\mathcal{P}}_{A} = |A\rangle\langle A|$, and $\Delta$ is the gap between them [Fig. \ref{Fig:JT}(b)]. 
We emphasize that the order of these vibronic states is predetermined by the geometric phase \cite{Ham1987} and does not depend on the computational methods [See \S 4.3 in Ref. \cite{Bersuker1989}] and not included weak interactions such as higher-order vibronic coupling, pseudo-JT coupling, and anharmonic terms [See Ref. \cite{Iwahara2018}].

{\it Vibronic quadrupole moment.---}
The vibronic states on the $d^1$ sites are correlated via intersite interactions acting on spin, orbital, and lattice degrees of freedom, underlying ordered phases of vibronic states \cite{Iwahara2018}. 
To characterize the phases, vibronic quadrupole operators are further introduced $\hat{\mathcal{T}}_\gamma$.
We define $\hat{\mathcal{T}}_\gamma$ by the projection of the electric quadrupole moments, $\tilde{\tau}_\gamma$, into the vibronic states, Eqs. (\ref{Eq:E}) and (\ref{Eq:A}): 
\begin{align}
 \hat{\mathcal{T}}_\gamma = \hat{\mathcal{P}} \tilde{\tau}_\gamma \hat{\mathcal{P}} \quad (\gamma = x,z),
 \label{Eq:T}
\end{align}
where $\hat{\mathcal{P}} = \sum_{\kappa = z^2, x^2-y^2, A} |\kappa\rangle \langle \kappa|$. 

Within the space of these vibronic states, the normal coordinates are proportional to $\hat{\mathcal{T}}_\gamma$. 
By the same procedure, we get 
\begin{align}
 q_{z^2(x^2-y^2)} \rightarrow \hat{\mathcal{P}} q_{z^2(x^2-y^2)} \hat{\mathcal{P}} 
 = -g \hat{\mathcal{T}}_{z(x)}.
 \label{Eq:q}
\end{align}
Thus the vibronic quadrupole moments enable a unified treatment of pseudo orbital and lattice degrees of freedom. 

From Eq. (\ref{Eq:q}), the thermal average of $\hat{\mathcal{T}}_\gamma$ is related to the deformation of the system. 
When $\mathcal{T}_z = \langle \hat{\mathcal{T}}_z \rangle > 0$ ($< 0$), the system is tetragonally compressed (elongated) along the $c$ axis. 
Similarly, $\mathcal{T}_x > 0$ ($< 0$) indicates the $-x^2+y^2$ ($x^2-y^2$) deformation within the $ab$ plane.

\begin{figure}[tb]
\centering
 \begin{tabular}{lllllll}
  \multicolumn{5}{l}{(a)} &~& (b) \\
  \multicolumn{5}{c}{
  \includegraphics[bb=0 0 720 255, height=0.22\linewidth]{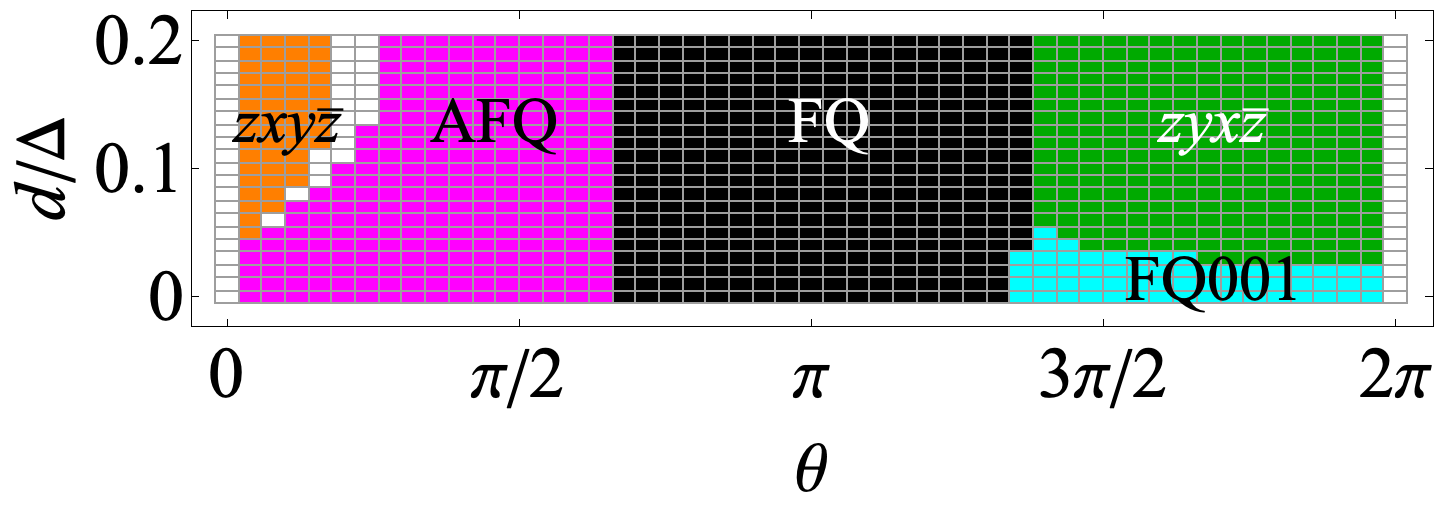}
  }
  & &
  \includegraphics[bb = 0 0 391 449, height=0.22\linewidth]{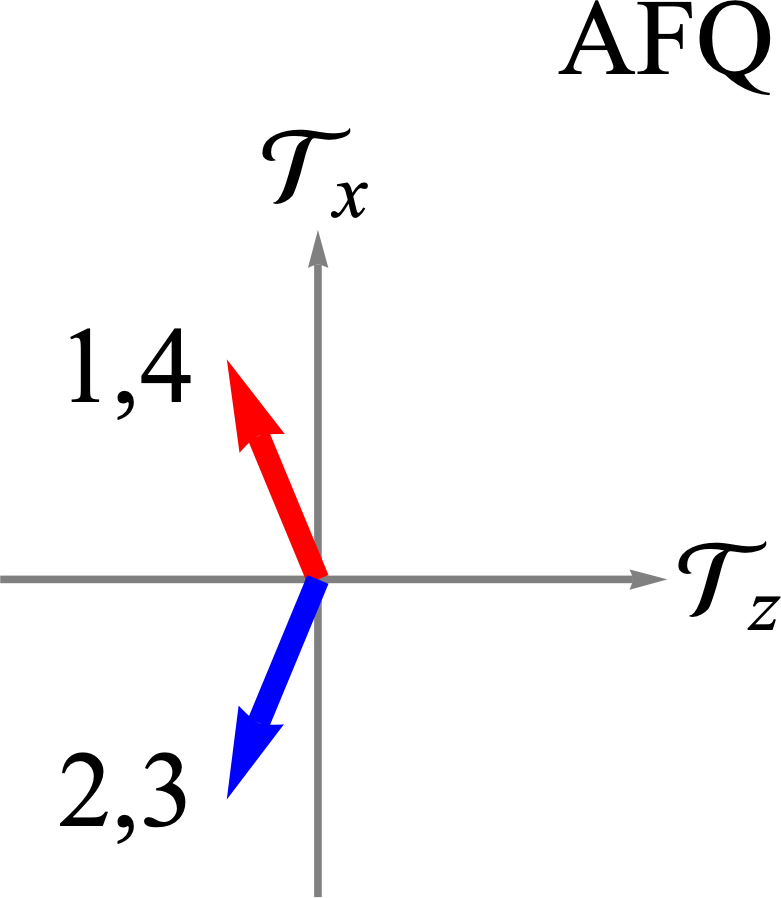}
  \\
  (c) &~& (d) &~& (e) &~& (f)\\
  \includegraphics[bb = 0 0 391 449, height=0.22\linewidth]{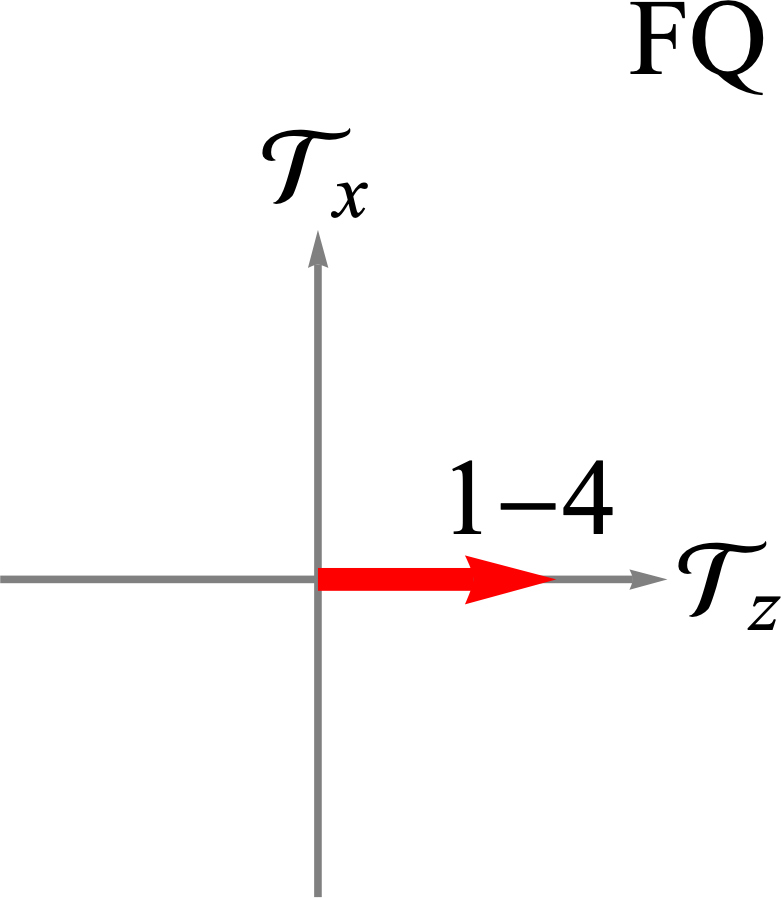}
  & &
  \includegraphics[bb = 0 0 391 449, height=0.22\linewidth]{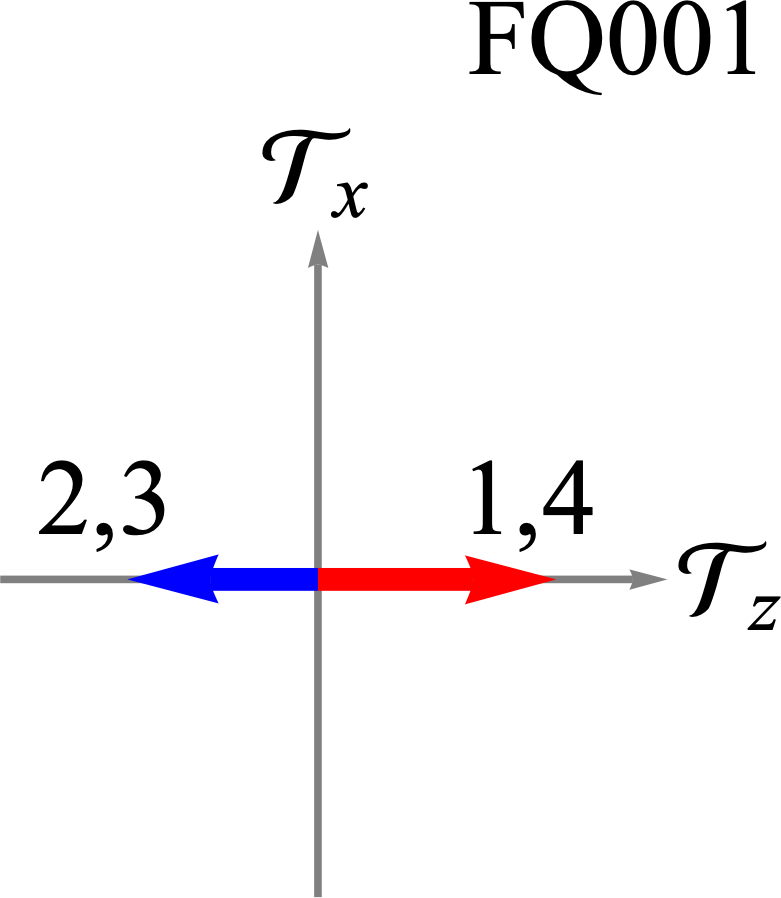}
  & &
  \includegraphics[bb = 0 0 391 449, height=0.22\linewidth]{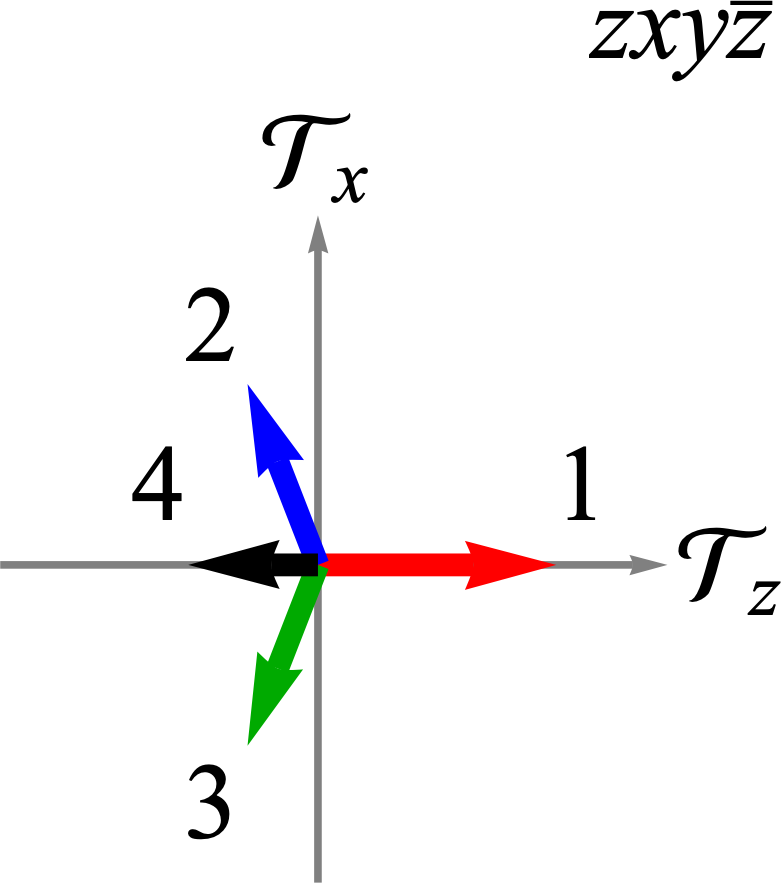}
  & &
  \includegraphics[bb = 0 0 391 449, height=0.22\linewidth]{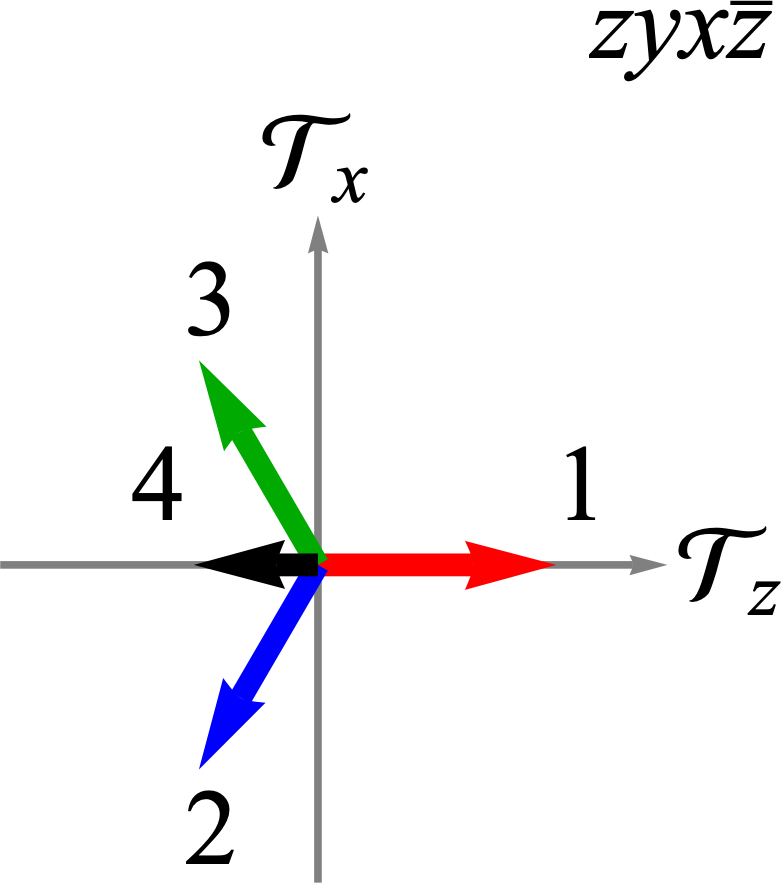}
\end{tabular}
 \caption{
 Vibronic ordered phases at $T=0$. 
 (a) Vibronic quadrupole phase diagram with respect to elastic coupling parameters, $d/\Delta$ and $\theta$.
  The magenta, black, cyan, green, orange, and white areas indicate AFQ, FQ, FQ001, AFM001, $zxy\bar{z}$, $zyx\bar{z}$, and intermediate phases, respectively. 
  (b)-(f) Arrangements of the vibronic quadrupole moments. 
  The numbers 1-4 indicate the metal sites in Fig. \ref{Fig:JT}(a).
 }
 \label{Fig:quadrupole}
\end{figure}

{\it Elastic quadrupole interaction.---}
For nearest neighbor sites $i,j$ on the $xy$ plane, we can write on symmetry reasons:
\begin{align}
\bm{D}_0^{ij}(\theta) = d_0 (\cos \theta \bm{\sigma}_0 + \sin \theta \bm{\sigma}_z).
\end{align}
Here $\bm{\sigma}_0$ is the 2-dimensional identity matrix, and $\bm{\sigma}_z$ are the $z$ component of the Pauli matrix.

The spin-orbit-lattice entangled states on the $d^1$ sites respond to the intersite elastic coupling.
Projecting $q_\alpha$ as in Eq. (\ref{Eq:q}),
$\hat{H}_\text{vib}$ reduces to a vibronic quadrupole interaction:
\begin{align}
 \hat{\mathcal{H}}^{ij}_\text{vib} = 
 (\hat{\mathcal{T}}_{z}^i, \hat{\mathcal{T}}_{x}^i)
 \bm{D}^{ij}(\theta)
 (\hat{\mathcal{T}}_{z}^j, \hat{\mathcal{T}}_{x}^j)^T,
 \label{Eq:Hel}
\end{align}
with $\bm{D}^{ij} = g^2 \bm{D}_0^{ij}$ and $d = g^2 d_0$. 
The present vibronic quadrupole model has the same mathematical form as the electric quadrupole model in Ref. \cite{Tsunetsugu2021}, while being of a completely different origin.

The vibronic quadrupole interaction (\ref{Eq:Hel}) with local vibronic Hamiltonian (\ref{Eq:H0}) shows various ferro- (FQ) and antiferro- (AFQ) vibronic quadrupole orders [Fig. \ref{Fig:quadrupole}(a)]. 
We first analyze the limiting case of $\Delta \rightarrow +\infty$ ($d/\Delta = 0$). 
To derive the ground state, we employ the four-sublattice mean-field theory as in Refs. \cite{Chen2010, Svoboda2021, Tsunetsugu2021}.
The $\mathcal{T}_x$ and $\mathcal{T}_z$ moments form the type-I AFM-like AFQ and FQ001 orders for $0 < \theta < \pi - \tan^{-1} 2$ and $\pi + \tan^{-1} 2 < \theta < 2\pi$, respectively [Fig. \ref{Fig:quadrupole} (b), (d)]. 
The FQ order of $\mathcal{T}_z$ $(>0, c/a < 1)$ arises for the other range of $\theta$ [Fig. \ref{Fig:quadrupole} (c)].

Reducing $\Delta$ ($d/\Delta > 0$), several phases emerge [Fig. \ref{Fig:quadrupole}(a)]. 
The FQ phase remains for almost the same range of $\theta$. 
The AFQ also persists, while for small $\theta \lesssim \pi/5$, a new phase ($zxy\bar{z}$) develops [Fig. \ref{Fig:quadrupole}(e)]. 
The FQ001 is fully quenched for $d/\Delta \gtrsim 0.05$ and a phase ($zyx\bar{z}$) [Fig. \ref{Fig:quadrupole}(f)] resembling to $zxy\bar{z}$ arises.

In the AFQ phase with finite $\Delta$, ferro $\mathcal{T}_z$ order ($\mathcal{T}_z < 0$, $c/a > 1$) develops too. 
When $\Delta$ is finite, $\hat{\mathcal{H}}_\text{vib}$ hybridizes the fully delocalized ground vibronic states and the excited one, giving rise to a tiny localization of the ground mean-field states around the minima of the APES. 
This localization makes $\mathcal{T}_z$ nonzero.

{\it Spin-orbital superexchange.---}
The pseudo spin and vibronic quadrupole moments on $d^1$ sites are correlated via spin-orbital superexchange interaction.
In a $xy$ ($yz, zx$) plane of the fcc lattice, the dominant electron transfer occurs between the nearest $xy$ ($yz, zx$) orbitals. 
We consider only the dominant electron transfer as in Ref. \cite{Romhanyi2017}.
Regarding the electron transfer interaction as a perturbation to the on-site Coulomb interaction, and applying the second-order perturbation theory, we obtain the spin-orbital superexchange model. 
Then, projecting the exchange coupling into the $\Gamma_8$ multiplets \cite{Romhanyi2017}, we finally get
\begin{align}
 \hat{H}_\text{ex}^{ij} &= \hat{J}^{ij} \tilde{\bm{s}}^i \cdot \tilde{\bm{s}}^j + \tilde{\bm{s}}^i \hat{\bm{K}}^{ij} \tilde{\bm{s}}^j + \hat{Q}^{ij}.
 \label{Eq:Hex1}
\end{align}
For a pair $i,j$ in a $xy$ plane, 
\begin{align}
\hat{J}^{ij} = \frac{8}{27}(2r_2+r_3) \left(\tilde{\tau}^i-\frac{1}{2}\right) \left(\tilde{\tau}^j-\frac{1}{2}\right), 
\label{Eq:J}
\end{align}
$\hat{\bm{K}}^{ij}$ takes a diagonal matrix form with 
\begin{align}
\hat{K}^{ij}_{xx} 
&= \frac{4}{3\sqrt{3}} (r_1-r_2) \left[\tilde{\tau}^i_x \left(\tilde{\tau}^j_z-\frac{1}{2}\right) + \left(\tilde{\tau}^i_z-\frac{1}{2}\right) \tilde{\tau}^j_x\right],
\nonumber\\
 \hat{K}^{ij}_{yy} &= -\hat{K}^{ij}_{xx},
\nonumber\\
 \hat{K}^{ij}_{zz} &= \frac{1}{4}(r_1-r_2)(2-\tilde{\tau}^i_z-\tilde{\tau}^j_z-4\tilde{\tau}^i_z\tilde{\tau}^j_z), 
\label{Eq:K}
\end{align}
and $\hat{Q}^{ij}$ is an electric quadrupole interaction 
\begin{align}
 \hat{Q}^{ij} = \frac{2}{27}(9r_1-r_2-2r_3)\tilde{\tau}^i_z\tilde{\tau}^j_z.
 \label{Eq:Q}
\end{align}
Here $r_1 = (1-3J_H/U)^{-1}$, $r_2 = (1-J_H/U)^{-1}$, and $r_3 = (1+2J_H/U)^{-1}$.
The units of $\hat{J}$, $\hat{\bm{K}}$, and $\hat{Q}$ are $J = t^2/U$.

The ground state of the exchange model (\ref{Eq:Hex1}) is either FM110 or coplanar AFM phase. 
Within the mean-field theory, the FM110 \cite{Chen2010} and coplanar AFM \cite{Romhanyi2017, Svoboda2021} phases arise for $1/3> J_H/U \gtrsim 0.23$ and the smaller $J_H/U$, respectively. 
Only the FM110 phase is relevant to the experimental data of $5d^1$ double perovskites.

We now include the vibronic effect into the exchange model, Eq. (\ref{Eq:Hex1}).
Projecting $\hat{H}_\text{ex}^{ij}$ into the space of the vibronic states by usings Eq. (\ref{Eq:T}), we obtain 
\begin{align}
 \hat{\mathcal{H}}_\text{ex}^{ij} &= \hat{\mathcal{J}}^{ij} \tilde{\bm{s}}^i \cdot \tilde{\bm{s}}^j + \tilde{\bm{s}}^i \hat{\mathcal{\bm{K}}}^{ij} \tilde{\bm{s}}^j + \hat{\mathcal{Q}}^{ij}.
 \label{Eq:Hex2}
\end{align}
Here $\hat{\mathcal{J}}^{ij}$, $\hat{\mathcal{\bm{K}}}^{ij}$, and $\hat{\mathcal{Q}}^{ij}$ are the exchange operators expressed via $\hat{\mathcal{T}}$, Eq. (\ref{Eq:T}).

\begin{figure}[tb]
\centering
 \begin{tabular}{lllllll}
 \multicolumn{5}{l}{(a)} &~& (b) \\
 \multicolumn{5}{c}{\includegraphics[bb=0 0 720 486, height=0.42\linewidth]{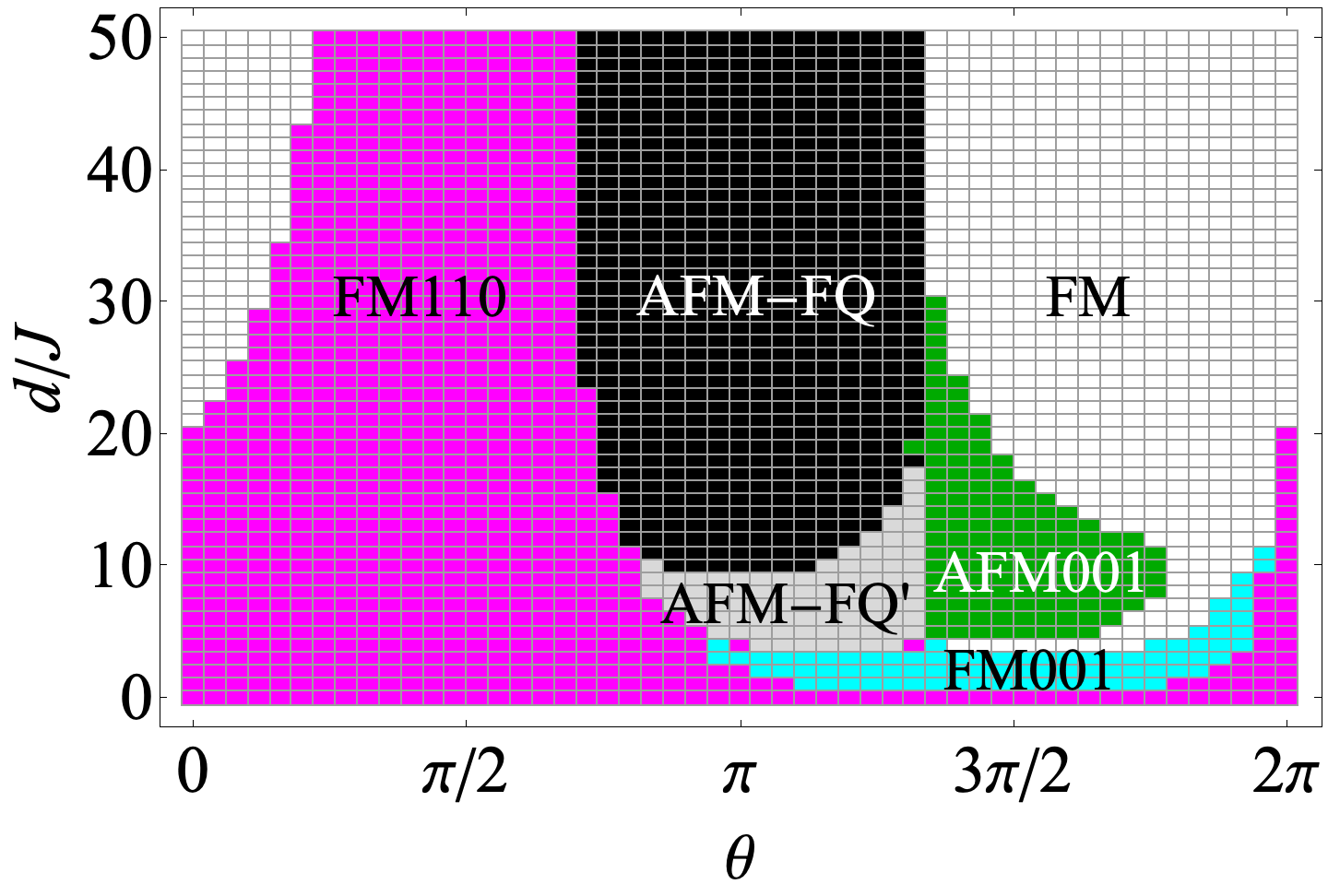}}
  & &
 \includegraphics[bb=0 0 400 805, height=0.42\linewidth]{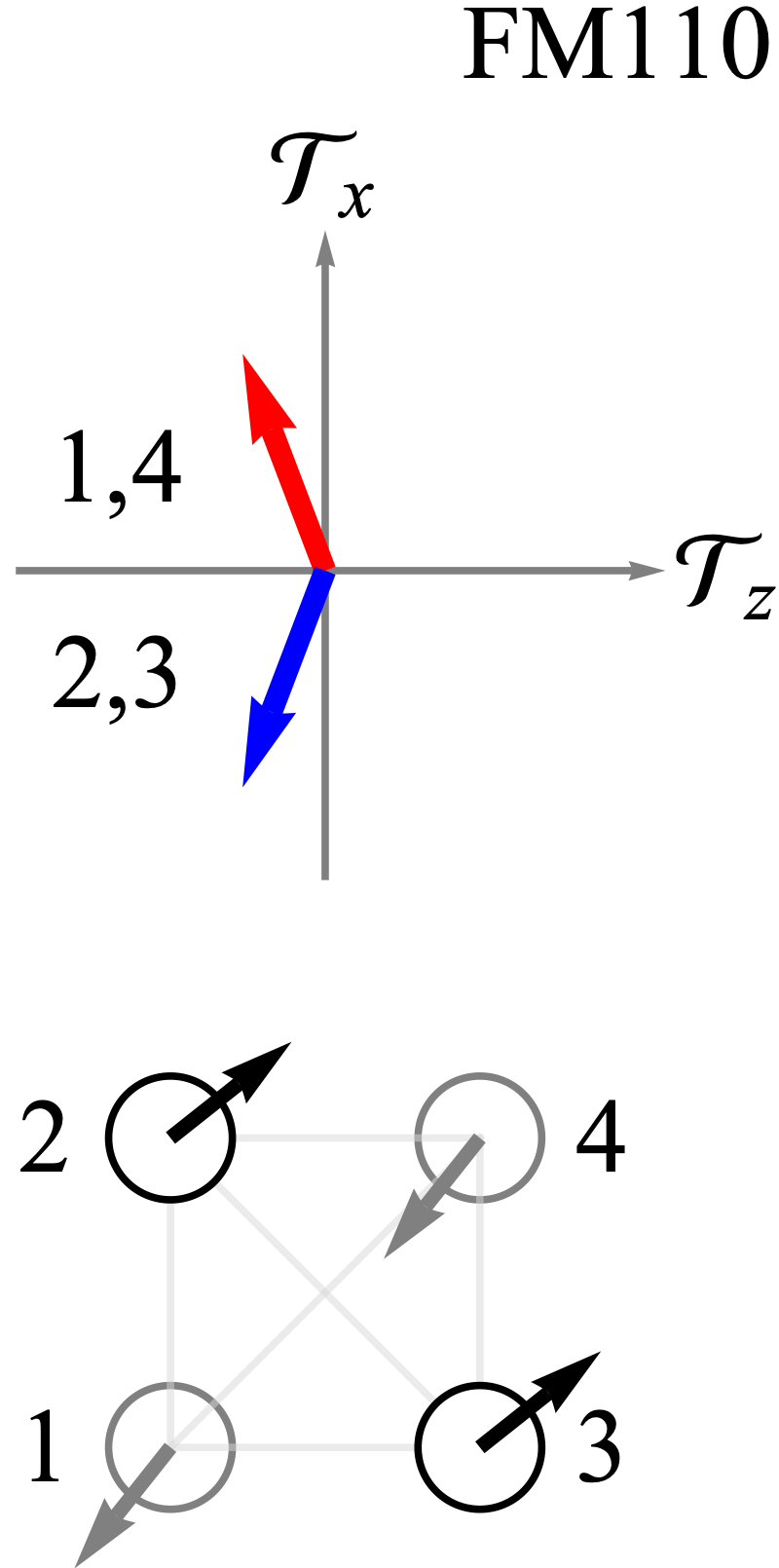}
  \\
 (c) &~& (d) &~& (e) &~& (f) \\ 
  \includegraphics[bb=0 0 400 805, height=0.42\linewidth]{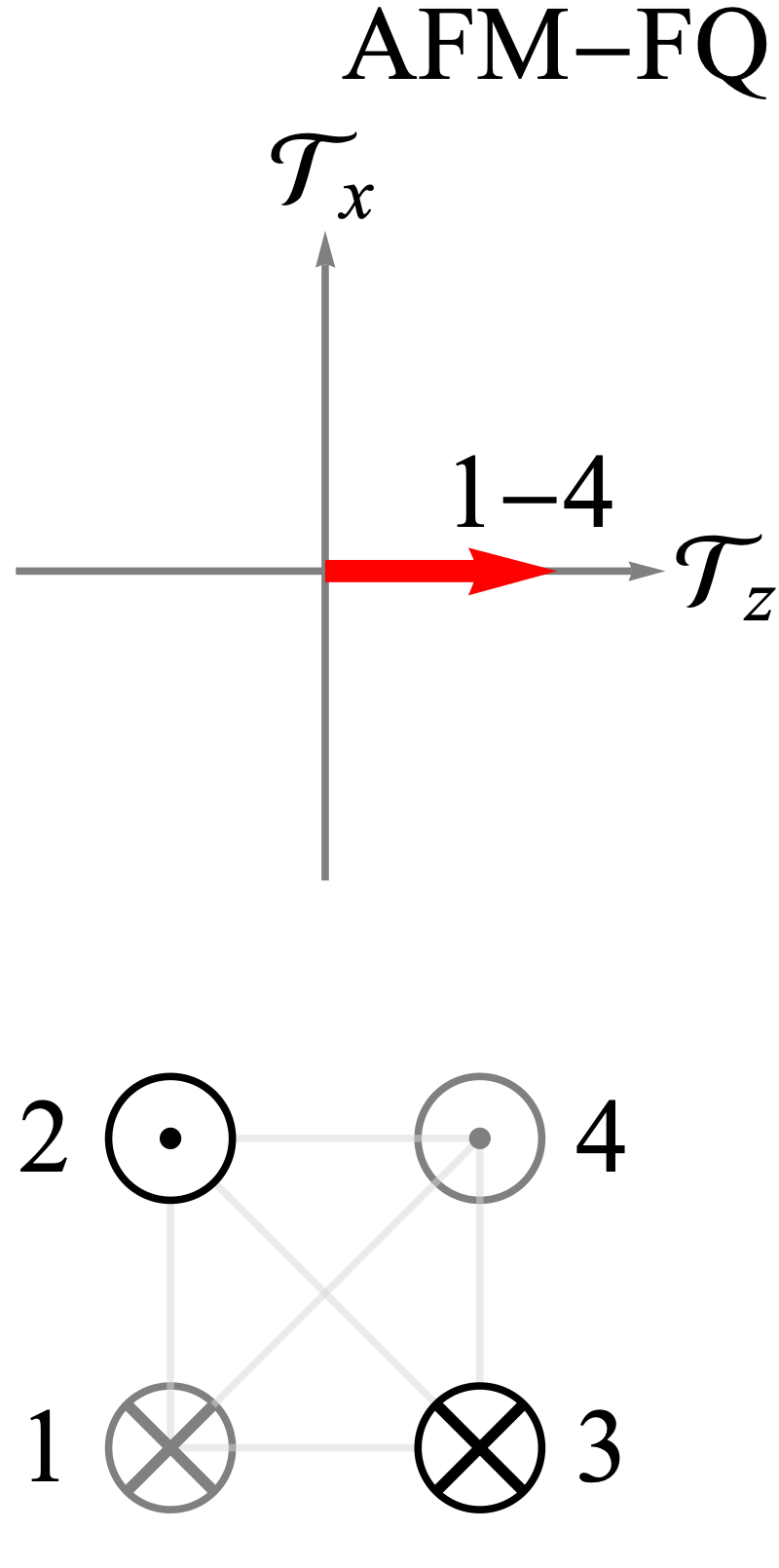}
  & &
  \includegraphics[bb=0 0 400 805, height=0.42\linewidth]{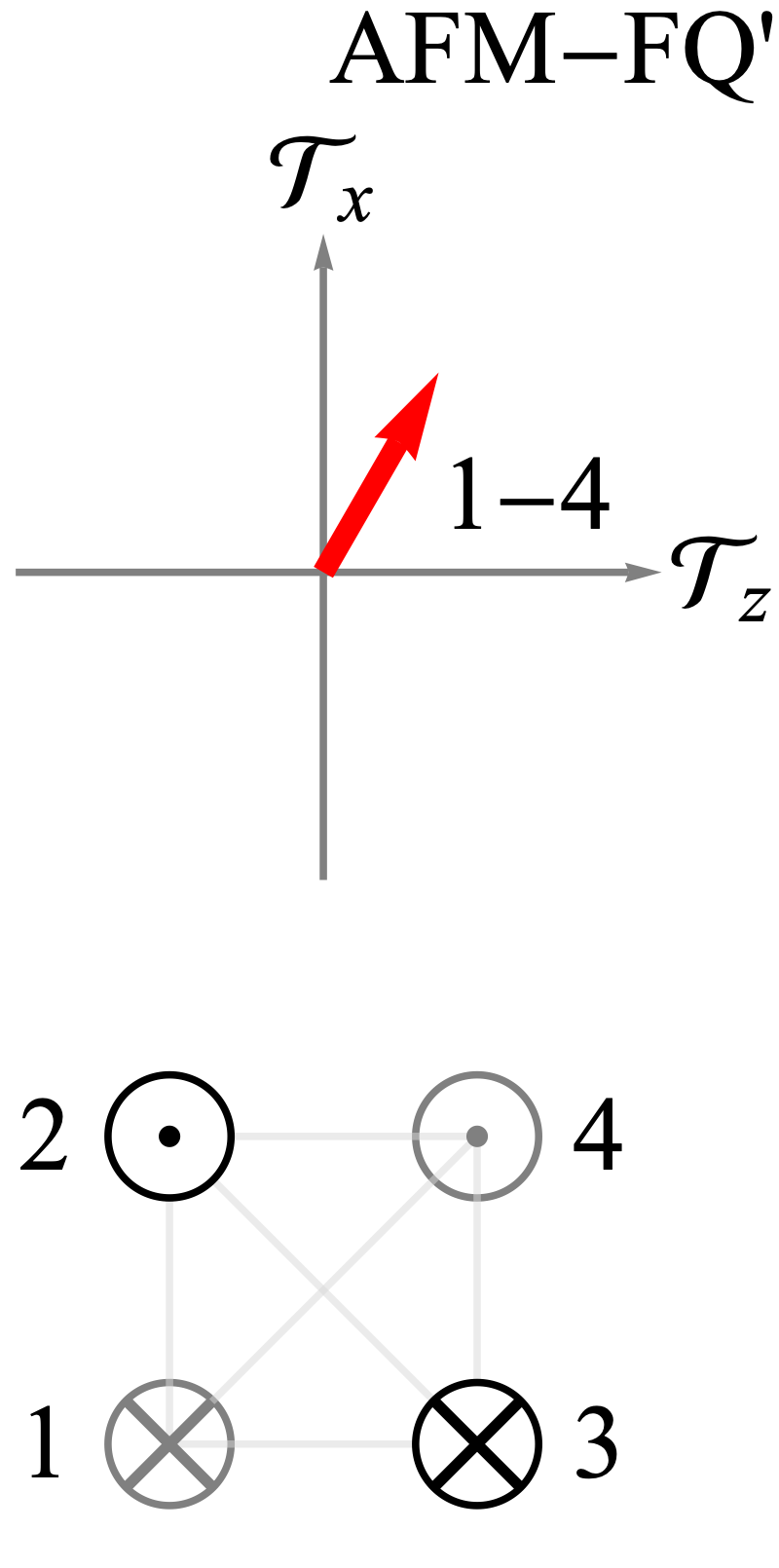}
  & &
  \includegraphics[bb=0 0 400 805, height=0.42\linewidth]{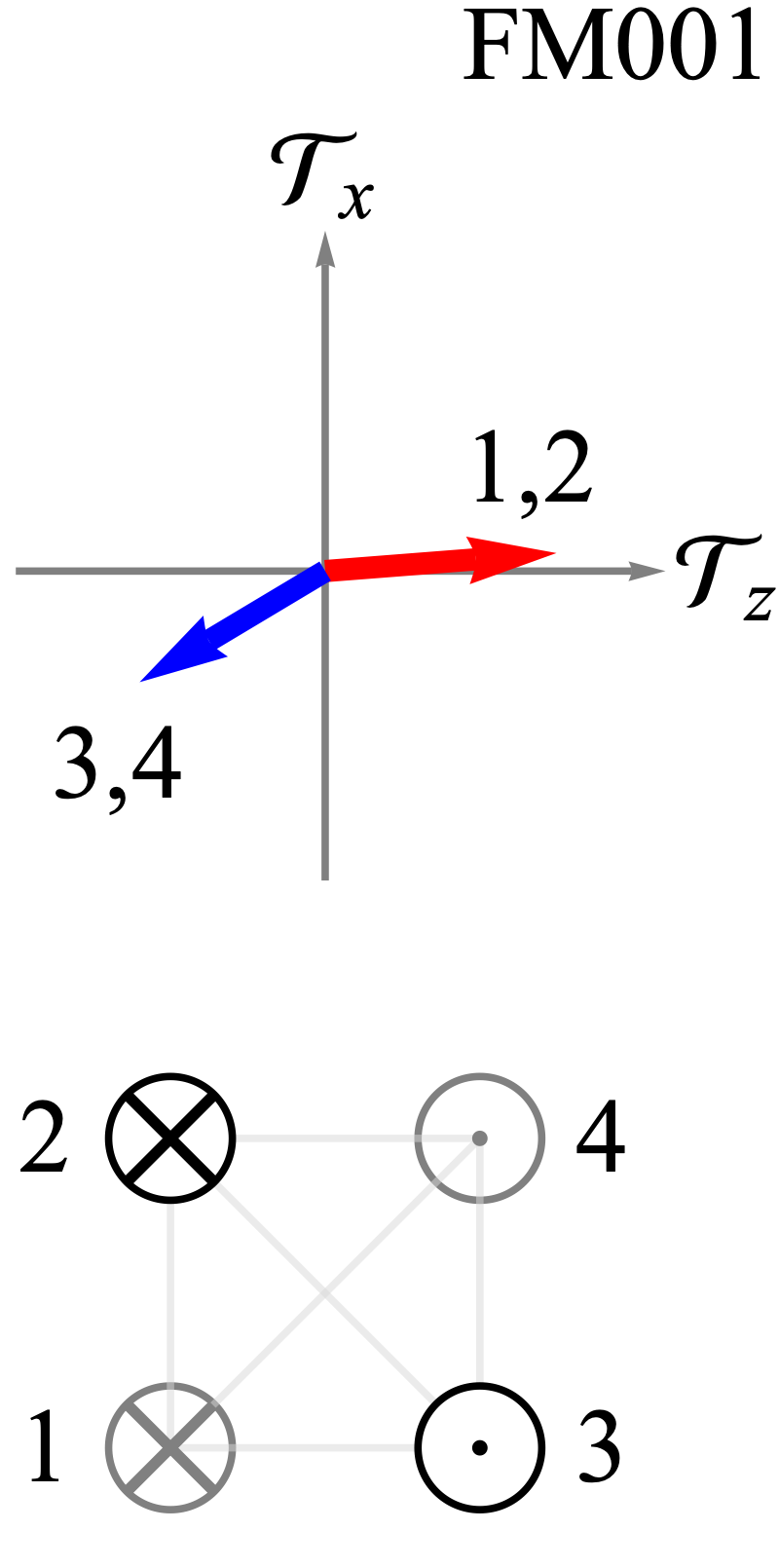}
  & &
  \includegraphics[bb=0 0 400 805, height=0.42\linewidth]{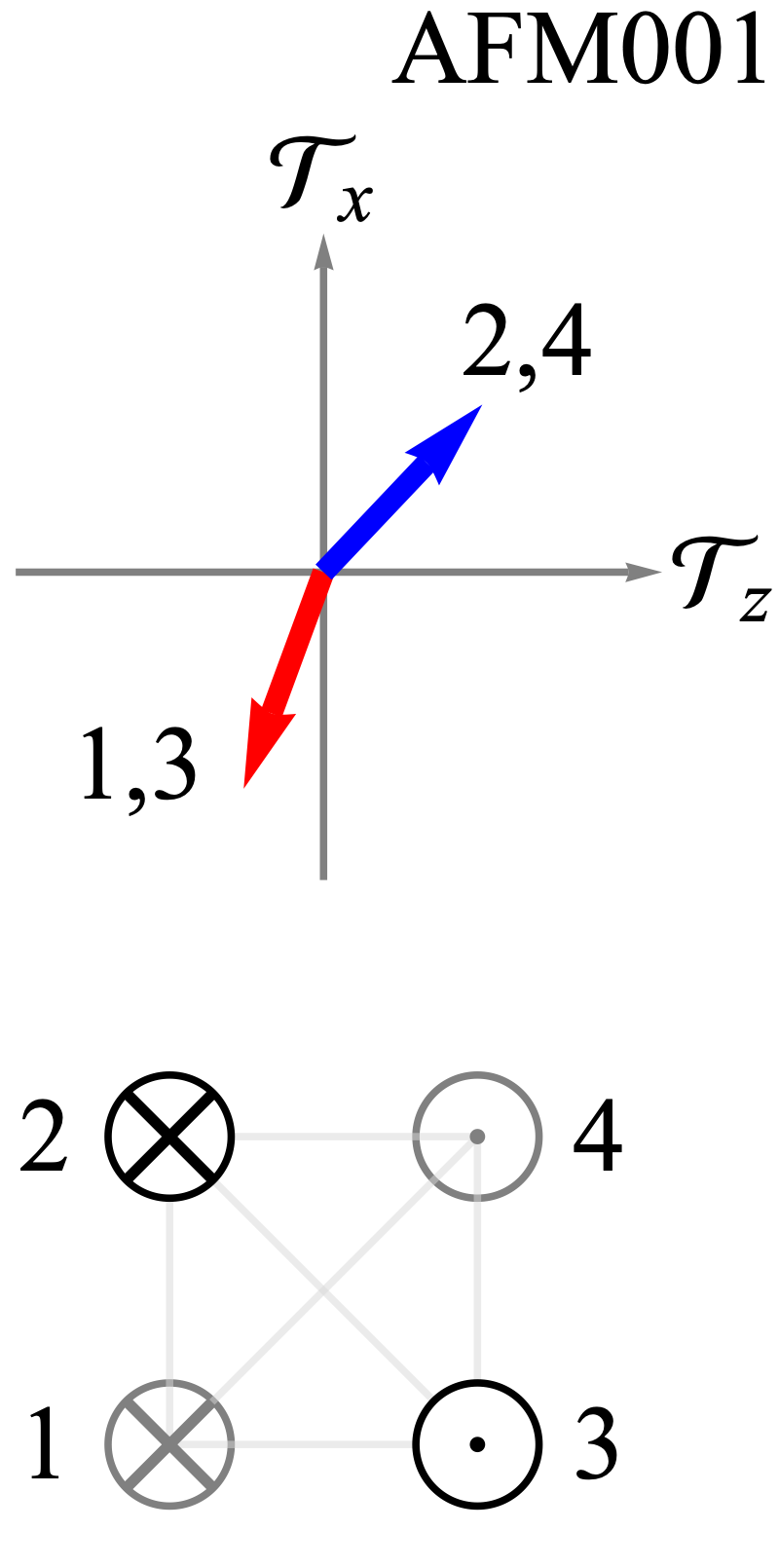}
  \end{tabular}
  \caption{
  Spin-vibronic ordered phases at $T = 0$.
  (a) Phase diagram with respect to $d/J$ and $\theta$.
  The magenta, black, gray, cyan, green, and white areas indicate FM110, AFM-FQ, AFM-FQ$^\prime$, FM001, AFM001, and non-symmetric FM phases, respectively.
  (b)-(f) Arrangements of the vibronic quadrupole moments (top) and pseudo spins from [001] direction (bottom). 
  See for sites 1-4 Fig. \ref{Fig:JT}(a).
  }
 \label{Fig:vibrospin}
\end{figure}

The spin-vibronic ordered phase is of FM110 type. 
The mean-field phase of $\hat{\mathcal{H}}_\text{ex}$ (\ref{Eq:Hex2}) with local $\hat{\mathcal{H}}_0$ (\ref{Eq:H0}) is either FM110-type or a two-sublattie ferromagnetic-type (FM001) [Fig. \ref{Fig:vibrospin}(b), (d)].
The FM110 appears for $\Delta > \Delta_c$ ($\Delta_c/J \approx 30$ for $J_H/U=0.3$) and $\Delta \lesssim 5J$, and the FM001 does otherwise. 
The dynamic JT effect would not drastically change the ordering of the ground state because $\Delta$ ($\approx$ several tens meV) is one or two orders of magnitude larger than $J$ ($\lesssim$ 1 meV) in $5d^1$ double perovskites \cite{Iwahara2018}.

{\it Vibronic order at 0 K.---}
Concomitantly treating local $\hat{\mathcal{H}}_0$, exchange $\hat{\mathcal{H}}_\text{ex}$, and elastic $\hat{\mathcal{H}}_\text{vib}$ couplings, we now analyze the ground ordered states at temperature $T=$ 0 K. 
We variationally derive the ground state within the mean-field theory.
Below we set $J_H/U = 0.3$ and $\Delta/J = 100$, with which the FM110 state appears when $\hat{\mathcal{H}}_\text{vib}=0$, and treat $d/J$ and $\theta$ as parameters.

Figure \ref{Fig:vibrospin}(a) shows the phase diagram with respect to $d$ and $\theta$. 
The FM110 phase persists in a similar range of $\theta$ where the AFQ order develops [Fig. \ref{Fig:quadrupole}(a)]. 
As in the AFQ phase with finite $\Delta$, $\hat{\mathcal{H}}_\text{vib}$ enhances the ferroquadrupole order of $\mathcal{T}_z < 0$ ($c/a > 1$). 

The FM110 phase turns into AFM phase by increasing $\theta$ in $\hat{\mathcal{H}}_\text{vib}$.
For sufficiently strong $\hat{\mathcal{H}}_\text{vib}$ ($d > d_c$. $d_c/J \approx 10$ in the present case), AFM-FQ phase emerges. 
In the AFM-FQ phase, the first kind AFM and ferro $\mathcal{T}_z$ ($>0$, $c/a < 1$) orders coexsist [Fig. \ref{Fig:vibrospin}(b)]. 
The strong $\hat{\mathcal{H}}_\text{vib}$ stabilizes the $|z^2, \tilde{s}_z\rangle$ spin-vibronic doublets, and the exchange interaction between the doublets on different sites becomes antiferromagnetic. 
For weaker $\hat{\mathcal{H}}_\text{vib}$ ($d_c > d > d_{c'}$. $d_{c'}/J \approx 4$ in the present case), AFM-FQ$^\prime$ phase appears. 
In the AFM-FQ$^\prime$ phase, the first kind AFM and a ferroic order of $\mathcal{T}$'s with $\mathcal{T}_x/\mathcal{T}_z = \sqrt{3}$ ($b = c$ and $a/c > 1$) or equivalent one develop [Fig. \ref{Fig:vibrospin}(c)]. 

We mainly focus on the FM110 and AFM-FQ phases because, as we discuss below, these phases show up in the existing materials.

Before we turn to the analysis of finite temperature phases, we discuss the effect of the electric quadrupole coupling.
The intersite Coulomb interaction contains a term acting on local electronic quadrupole moments $\tilde{\tau}_\gamma$ \cite{Chen2010}. 
Within the space of the vibronic states, the electric quadrupole interaction takes the same form as $\hat{\mathcal{H}}_\text{vib}$ (\ref{Eq:Hel}) with fixed $\theta = 0.587 \pi$. 
Figure \ref{Fig:vibrospin}(a) shows that the electric quadrupole interaction does not qualitatively change the FM110 phase for arbitrary strength of the interaction.

\begin{figure}[bt]
\centering
 \begin{tabular}{lll}
 (a) &~& (b) \\
 \includegraphics[bb = 0 0 720 613, height=0.41\linewidth]{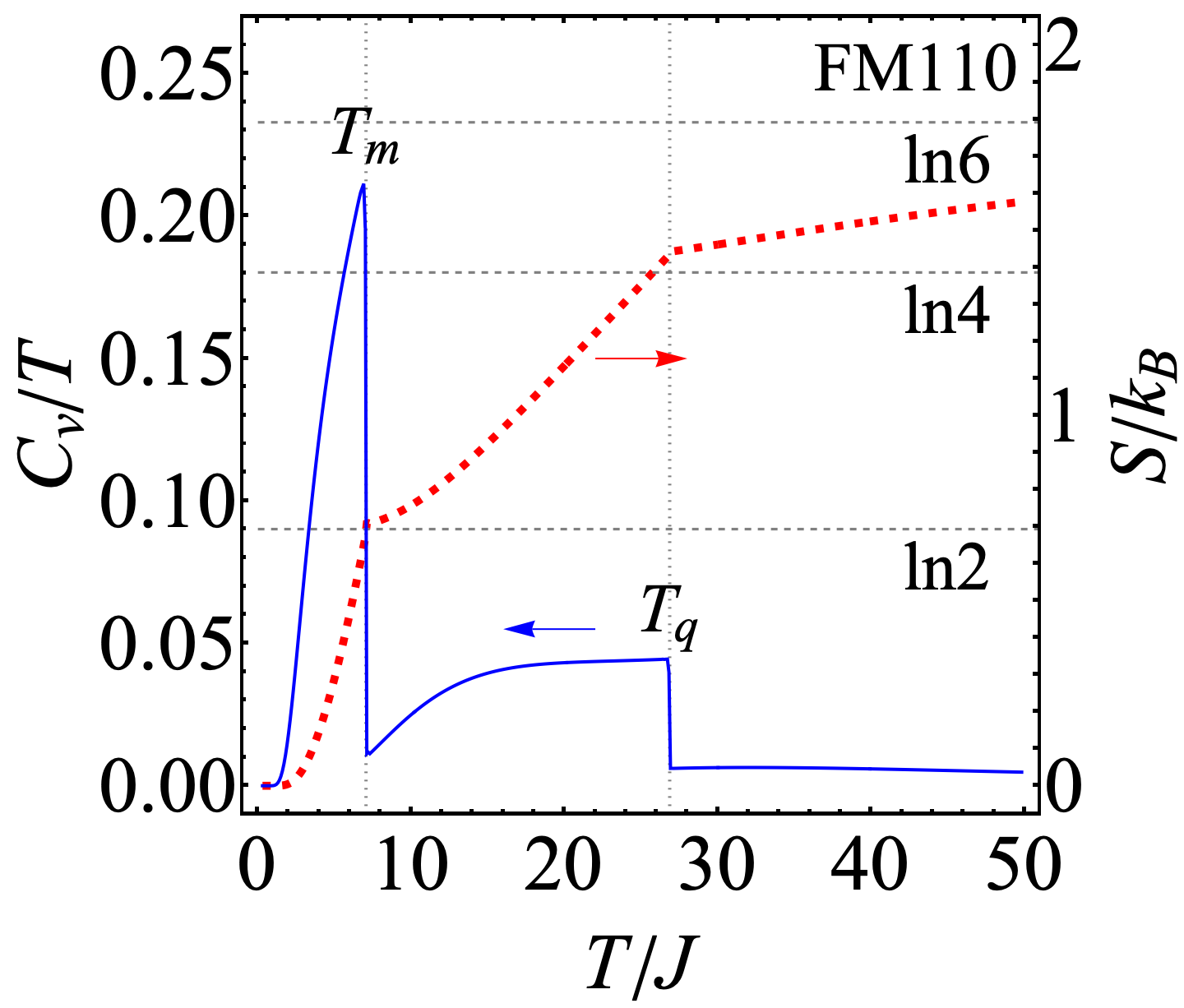}
 & &
 \includegraphics[bb = 0 0 720 659, height=0.41\linewidth]{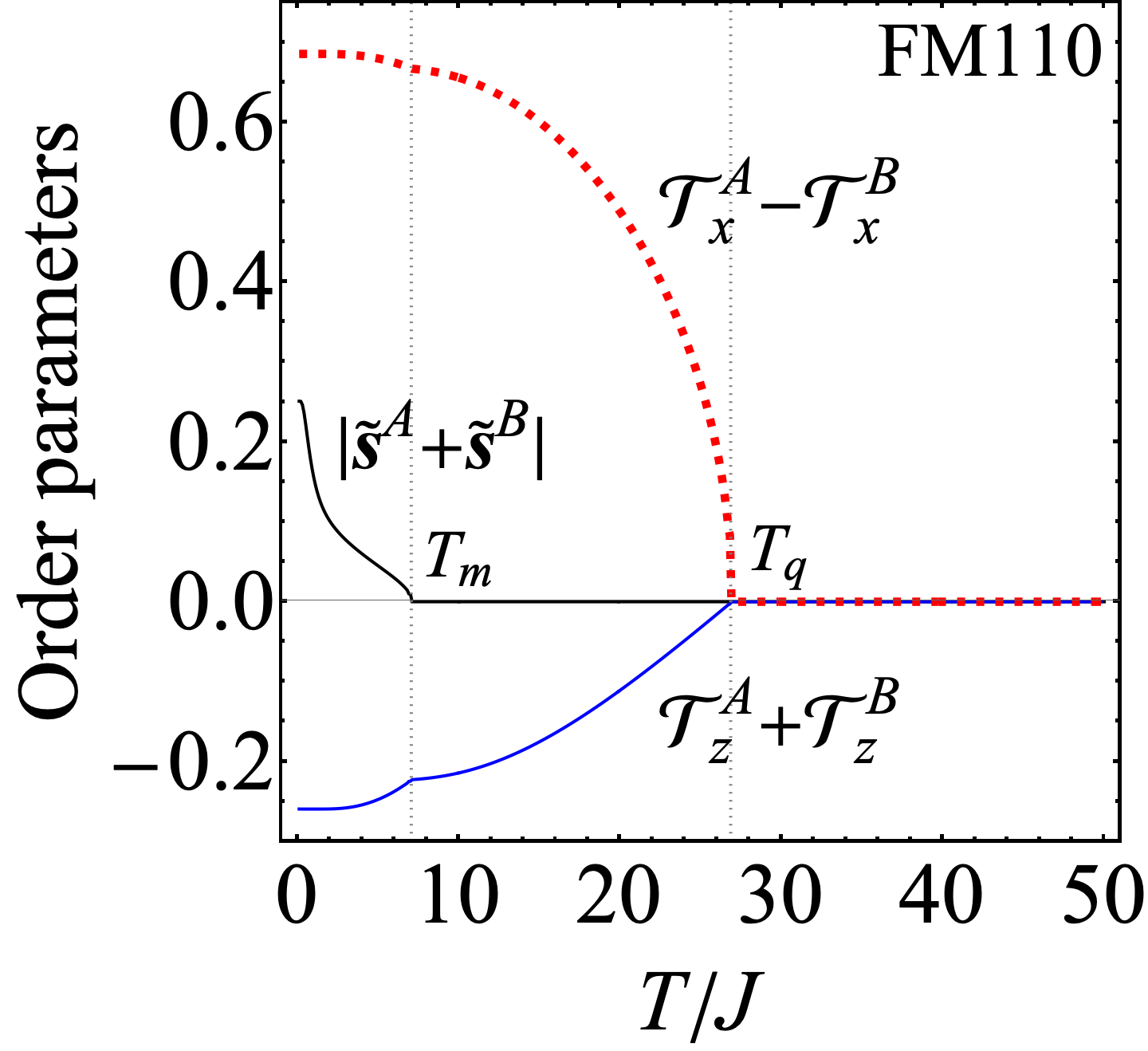}
 \\
 (c) &~& (d) \\
 \includegraphics[bb = 0 0 720 613, height=0.41\linewidth]{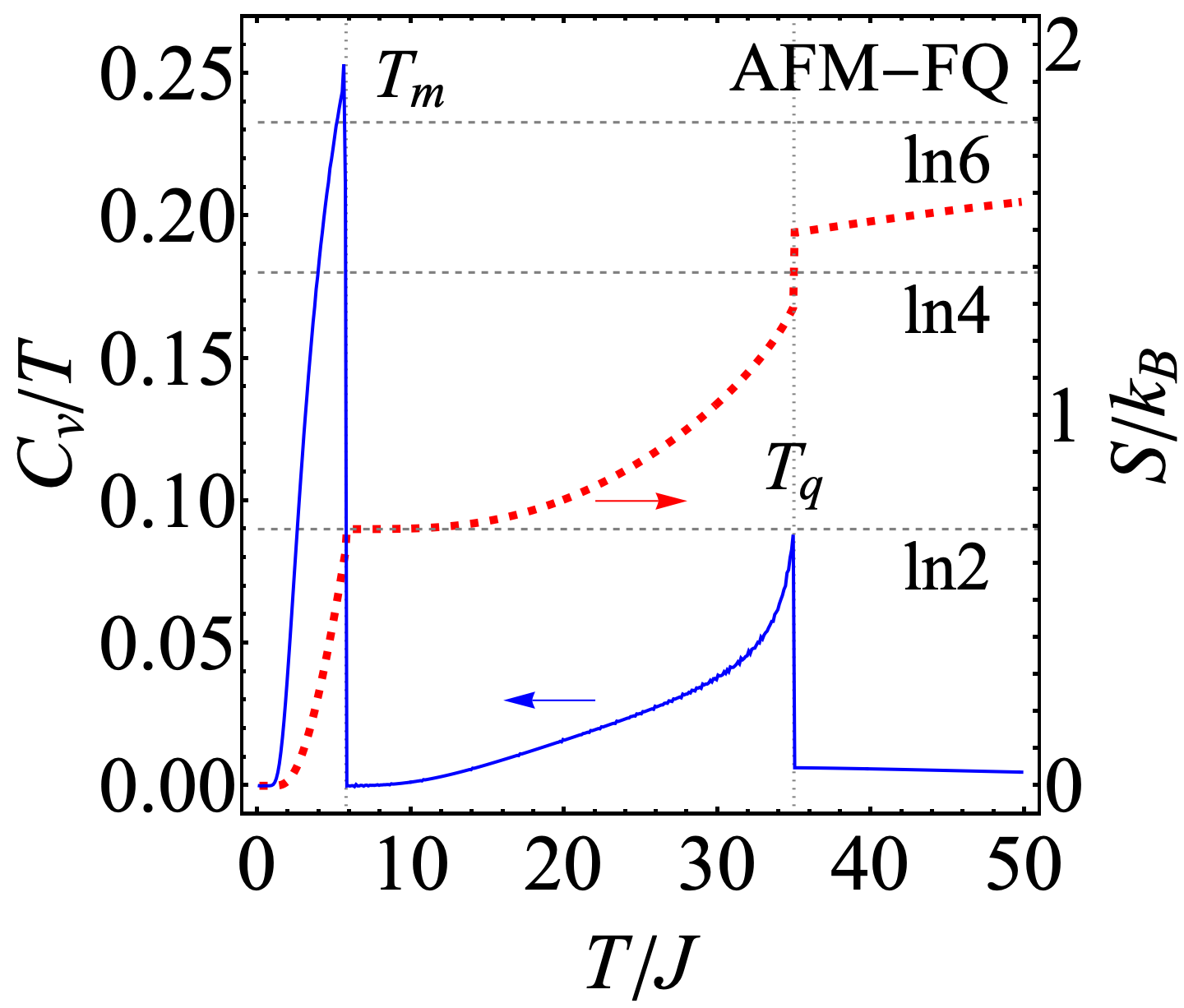}
 & &
 \includegraphics[bb = 0 0 720 716, height=0.41\linewidth]{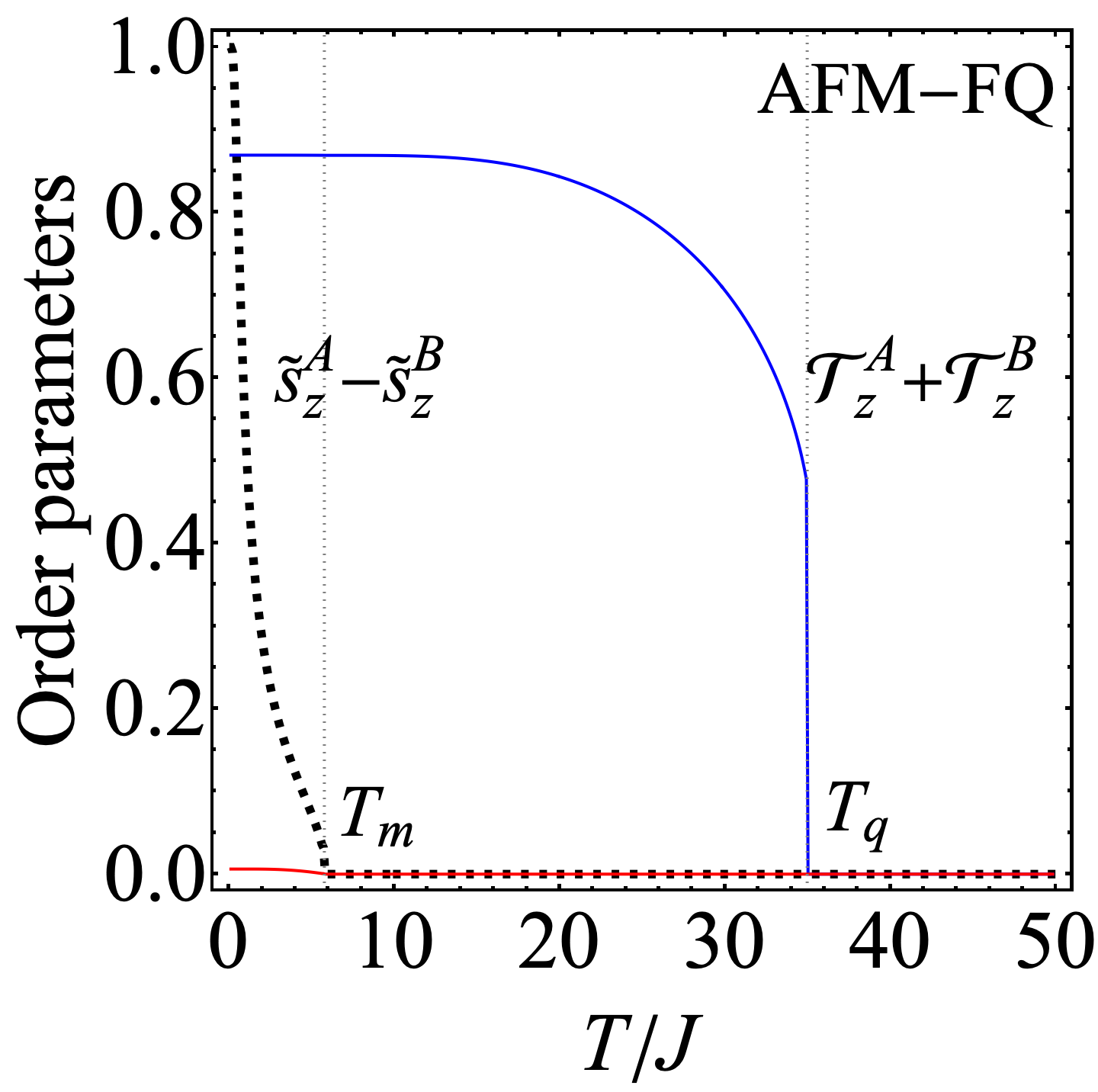}
 \end{tabular}
 \caption{
  Temperature evolution of the order parameters and thermodynamic quantities in the FM110 and AFM-FQ phases.  
  In (a), (c), the blue solid and red dotted lines indicate $C_v$ and $S$, respectively. 
  In (b), (d), the red and blue lines indicate vibronic quadrupole moments and the black lines pseudo spins. 
  The FM110 and AFM-FQ phases have two sublattices, $A$ and $B$. 
  $\phi = 12\pi/25$ for FM110 phase and $\pi$ for the AFM-FQ phase. 
 }
 \label{Fig:OP}
\end{figure}

{\it Finite temperature vibronic order.---}
Let us now analyze the temperature evolutions of the FM110 and AFM-FQ phases. 
For $d > d_c$, in both states, the specific heat $C_v$ shows two phase transitions at $T_m$ and $T_q$ ($T_m < T_q$) [Fig. \ref{Fig:OP}(a), (c)]. 
The order parameters indicate that the transitions at $T_m$ and $T_q$ correspond to the magnetic and vibronic quadrupole transitions, respectively [Fig. \ref{Fig:OP}(b), (d)]. 
At these transitions, the entropy gains the contributions ($k_B \text{ln}2$ each) from the magnetic and the ground vibronic degrees of freedom, respectively [Fig. \ref{Fig:OP}(a), (c)]. 
Decreasing $d$, $T_q$ approaches $T_m$, and below $d_c$, the two transitions merge.

The presence or absence of the vibronic quadrupole transition depends on the strength of the correlation between $\tilde{\bm{s}}$ and $\hat{\mathcal{T}}$ via $\hat{\mathcal{H}}_\text{ex}$ [See the first and second terms in Eq. (\ref{Eq:Hex2})].
When $\hat{\mathcal{H}}_\text{vib}$ is dominant ($d > d_c$), $\hat{\mathcal{T}}$'s are almost independent from $\tilde{\bm{s}}$. 
Under this situation, $T_q$ (or $T_q/T_m$) and $\mathcal{T}$ become larger as $d$ increases. 
When $\hat{\mathcal{H}}_\text{ex}$ is dominant ($d \leq d_c$), the pseudo spin and vibronic states are strongly correlated, and only one transition appears.

Finally, we note that the ferro $\mathcal{T}_z$ order in the FM110 phase persists above $T_m$ ($d > d_c$) [Fig. \ref{Fig:OP}(b)].
The ferro $\mathcal{T}_z$ order originates from the hybridization of the $E$ and $A$ vibronic states by $\hat{\mathcal{H}}_\text{vib}$ [Fig. \ref{Fig:JT} (b)] as in the AFQ phase [Fig. \ref{Fig:quadrupole}(c)].

{\it Vibronic order in materials.---}
Our spin-vibronic orders are consistent with the ordered phases of the $5d^1$ double perovskites.
Particularly, the high-temperature quadrupole phase of Ba$_2$MgReO$_6$ and the AFM phase of $A_2$Ta$X_6$.

The theoretical quadrupole phase above $T_m$ of the FM110 phase [Fig. \ref{Fig:OP}(a), (b)] explains the experimental data of Ba$_2$MgReO$_6$ \cite{Hirai2020}.
In the rhenium compound, the ferro $\mathcal{T}_z$ ($<0$, $c/a>1$) and antiferro $\mathcal{T}_x$ quadrupole orders coexist between $T_m = 18$ K and $T_q = 33$ K, which agrees with our calculation 
[Fig. \ref{Fig:OP}(b)].
The specific heat of the single-crystalline sample does not show a clear peak \cite{Hirai2019}, which agrees with the present simulation [Fig. \ref{Fig:OP}(a)].

In similar FM110 compounds such as Ba$_2$NaOsO$_6$ and Ba$_2$ZnReO$_6$, the $\mathcal{T}_z$ moment has not been experimentally detected, which does not contradict our theory.  
As mentioned above, $T_q/T_m$ and $\mathcal{T}_z$ increase as $\hat{\mathcal{H}}_\text{vib}$ becomes stronger. 
Indeed, $T_q/T_m$'s are about 1.3 for Ba$_2$NaOsO$_6$ \cite{Willa2019} and 1.4 for Ba$_2$ZnReO$_6$ \cite{Woodward2022}, and smaller than $T_q/T_m \approx$ 1.8 for Ba$_2$MgReO$_6$, indicating that $\hat{\mathcal{H}}_\text{vib}$'s (and $\mathcal{T}_z$) of the former compounds are weaker than that of the latter. 
$\mathcal{T}_z$ of the former could be too small to experimentally detect with currently available methods.

Our AFM-FQ phase [Fig. \ref{Fig:vibrospin}(a), (c)] captures the low-temperature ordered phases of $A_2$Ta$X_6$.
Lowering $T$ from the room temperature, 
the tantalum compounds undergo tetragonal compression at $T_q$ ($c/a<1$, $\mathcal{T}_z>0$), continuing through the N\'{e}el transition \cite{Ishikawa2019, Ishikawa2021, Tehrani2023}, which agrees with the temperature evolution of our AFM-FQ phase [Fig. \ref{Fig:OP}(d)].
At these transitions, the specific heat of single-crystalline Cs$_2$TaCl$_6$ shows sharp peaks \cite{Tehrani2023}, which is consistent with our calculation [Fig. \ref{Fig:OP}(c)].
The AFM-FQ phase is also in line with the experimental data of Ba$_2$CaReO$_6$ \cite{Yamaura2006, Ishikawa2021b}.

The AFM-FQ phase could appear in cubic Ba$_2$LiOsO$_6$ too. 
Although the presence of the $\mathcal{T}_z$ quadrupole order has not been reported yet \cite{Stitzer2002, Steele2011, Woodward2022}, 
the magnitude of $\mathcal{T}_z$ could be too small to experimentally detect because of weak $\hat{\mathcal{H}}_\text{vib}$ as in isostructural and isoelectronic Ba$_2$NaOsO$_6$. 
Recent NMR data (the temperature evolution of $1/T_2$) show a bump just above $T_m$ \cite{Cong2022}. 
As discussed in the literature, the data might imply the presence of quadrupole transition.

Furthermore, the FM110-to-AFM transition resolves why similar $5d^1$ double perovskites exhibit FM110 and AFM phases. 
Isoelectronic and isostructural Ba$_2$NaOsO$_6$ and Ba$_2$LiOsO$_6$ show FM110 and AFM phases, respectively. 
The nature of the phase of Ba$_2$CdReO$_6$ is under debate whether it is FM110 \cite{Hirai2021} or AFM \cite{Woodward2022}.
We propose that these systems are close to the border of the FM110 and AFM phases in the phase diagram [Fig. \ref{Fig:vibrospin}(a)]: by slightly varying $\theta$ in $\hat{\mathcal{H}}_\text{vib}$, phase transition between them could occur. 
The elastic coupling varies by changing nonmagnetic alkali ions in the osmium compounds and via a small change in, e.g., crystal anisotropy and the density of defects of different samples of the rhenium compound.

{\it Conclusion.---}
We developed a microscopic vibronic approach that concomitantly treats the competing spin-orbit and vibronic interactions in cubic $d^1$ double perovskites. 
We found that the magnetic order coexists with the vibronic order characterized by ferro/antiferro arrangement of vibronic quadrupole moments on sites. 
The present theory allows the rationalization of the mechanism of puzzling phases in the $5d^1$ double perovskites:
the high-temperature quadrupole orders above the magnetic transition of FM110 phase in rhenium compounds and the antiferromagnetic phase with tetragonal compression in tantalum compounds.

The vibronic quadrupole order can be detected by resonant x-ray scattering. 
Since the vibronic quadrupole moments, Eq. (\ref{Eq:T}), respond to all external fields acting on the orbitals, high-resolution resonant elastic/inelastic x-ray scattering measurements used to probe orbital orders \cite{Ishihara2002, Fabrizio1998, Ament2011} will be indispensable to study vibronic order too. 
A fingerprint of the vibronic order will appear as the anisotropy of the atomic scattering factors and in the low-energy vibronic excitation bands in resonant inelastic x-ray scattering spectra. 

The unquenched orbital-lattice entanglement could exist in a large class of metal compounds. 
The proposed approach is applicable to systems with orbital (quasi) degeneracy on sites since the orbital-lattice entanglement could persist in low-symmetric systems with moderate orbital energy splitting. 
The dynamic vibronic order assumes the formation of local vibronic states: this picture is adequate in the systems where intersite interactions are weak and do not intermix many vibronic states on sites. 
Further investigations on both experimental and theoretical sides are required to get deeper insights into the nature of vibronic order.

\begin{acknowledgments}
This work was partly supported by Okura Kazuchika memorial foundation and Grant-in-Aid for Scientific Research (Grant No. 22K03507) from the Japan Society for the Promotion of Science.
\end{acknowledgments}


\begin{thebibliography}{40}%
\makeatletter
\providecommand \@ifxundefined [1]{%
 \@ifx{#1\undefined}
}%
\providecommand \@ifnum [1]{%
 \ifnum #1\expandafter \@firstoftwo
 \else \expandafter \@secondoftwo
 \fi
}%
\providecommand \@ifx [1]{%
 \ifx #1\expandafter \@firstoftwo
 \else \expandafter \@secondoftwo
 \fi
}%
\providecommand \natexlab [1]{#1}%
\providecommand \enquote  [1]{``#1''}%
\providecommand \bibnamefont  [1]{#1}%
\providecommand \bibfnamefont [1]{#1}%
\providecommand \citenamefont [1]{#1}%
\providecommand \href@noop [0]{\@secondoftwo}%
\providecommand \href [0]{\begingroup \@sanitize@url \@href}%
\providecommand \@href[1]{\@@startlink{#1}\@@href}%
\providecommand \@@href[1]{\endgroup#1\@@endlink}%
\providecommand \@sanitize@url [0]{\catcode `\\12\catcode `\$12\catcode
  `\&12\catcode `\#12\catcode `\^12\catcode `\_12\catcode `\%12\relax}%
\providecommand \@@startlink[1]{}%
\providecommand \@@endlink[0]{}%
\providecommand \url  [0]{\begingroup\@sanitize@url \@url }%
\providecommand \@url [1]{\endgroup\@href {#1}{\urlprefix }}%
\providecommand \urlprefix  [0]{URL }%
\providecommand \Eprint [0]{\href }%
\providecommand \doibase [0]{https://doi.org/}%
\providecommand \selectlanguage [0]{\@gobble}%
\providecommand \bibinfo  [0]{\@secondoftwo}%
\providecommand \bibfield  [0]{\@secondoftwo}%
\providecommand \translation [1]{[#1]}%
\providecommand \BibitemOpen [0]{}%
\providecommand \bibitemStop [0]{}%
\providecommand \bibitemNoStop [0]{.\EOS\space}%
\providecommand \EOS [0]{\spacefactor3000\relax}%
\providecommand \BibitemShut  [1]{\csname bibitem#1\endcsname}%
\let\auto@bib@innerbib\@empty
\bibitem [{\citenamefont {Rau}\ \emph {et~al.}(2016)\citenamefont {Rau},
  \citenamefont {Lee},\ and\ \citenamefont {Kee}}]{Rau2016}%
  \BibitemOpen
  \bibfield  {author} {\bibinfo {author} {\bibfnamefont {J.~G.}\ \bibnamefont
  {Rau}}, \bibinfo {author} {\bibfnamefont {E.~K.-H.}\ \bibnamefont {Lee}},\
  and\ \bibinfo {author} {\bibfnamefont {H.-Y.}\ \bibnamefont {Kee}},\
  }\bibfield  {title} {\bibinfo {title} {{Spin-Orbit Physics Giving Rise to
  Novel Phases in Correlated Systems: Iridates and Related Materials}},\ }\href
  {https://doi.org/10.1146/annurev-conmatphys-031115-011319} {\bibfield
  {journal} {\bibinfo  {journal} {Annu. Rev. Condens. Matter Phys.}\ }\textbf
  {\bibinfo {volume} {7}},\ \bibinfo {pages} {195} (\bibinfo {year}
  {2016})}\BibitemShut {NoStop}%
\bibitem [{\citenamefont {Takagi}\ \emph {et~al.}(2019)\citenamefont {Takagi},
  \citenamefont {Takayama}, \citenamefont {Jackeli}, \citenamefont
  {Khaliullin},\ and\ \citenamefont {Nagler}}]{Takagi2019}%
  \BibitemOpen
  \bibfield  {author} {\bibinfo {author} {\bibfnamefont {H.}~\bibnamefont
  {Takagi}}, \bibinfo {author} {\bibfnamefont {T.}~\bibnamefont {Takayama}},
  \bibinfo {author} {\bibfnamefont {G.}~\bibnamefont {Jackeli}}, \bibinfo
  {author} {\bibfnamefont {G.}~\bibnamefont {Khaliullin}},\ and\ \bibinfo
  {author} {\bibfnamefont {S.~E.}\ \bibnamefont {Nagler}},\ }\bibfield  {title}
  {\bibinfo {title} {{Concept and realization of Kitaev quantum spin
  liquids}},\ }\href {https://doi.org/10.1038/s42254-019-0038-2} {\bibfield
  {journal} {\bibinfo  {journal} {Nat. Rev. Phys.}\ }\textbf {\bibinfo {volume}
  {1}},\ \bibinfo {pages} {264} (\bibinfo {year} {2019})}\BibitemShut {NoStop}%
\bibitem [{\citenamefont {Stitzer}\ \emph {et~al.}(2002)\citenamefont
  {Stitzer}, \citenamefont {Smith},\ and\ \citenamefont {{zur
  Loye}}}]{Stitzer2002}%
  \BibitemOpen
  \bibfield  {author} {\bibinfo {author} {\bibfnamefont {K.~E.}\ \bibnamefont
  {Stitzer}}, \bibinfo {author} {\bibfnamefont {M.~D.}\ \bibnamefont {Smith}},\
  and\ \bibinfo {author} {\bibfnamefont {H.-C.}\ \bibnamefont {{zur Loye}}},\
  }\bibfield  {title} {\bibinfo {title} {{Crystal growth of Ba$_2M$OsO$_6$
  ($M=$ Li, Na) from reactive hydroxide fluxes}},\ }\href
  {https://doi.org/https://doi.org/10.1016/S1293-2558(01)01257-2} {\bibfield
  {journal} {\bibinfo  {journal} {Solid State Sci.}\ }\textbf {\bibinfo
  {volume} {4}},\ \bibinfo {pages} {311} (\bibinfo {year} {2002})}\BibitemShut
  {NoStop}%
\bibitem [{\citenamefont {Yamamura}\ \emph {et~al.}(2006)\citenamefont
  {Yamamura}, \citenamefont {Wakeshima},\ and\ \citenamefont
  {Hinatsu}}]{Yamaura2006}%
  \BibitemOpen
  \bibfield  {author} {\bibinfo {author} {\bibfnamefont {K.}~\bibnamefont
  {Yamamura}}, \bibinfo {author} {\bibfnamefont {M.}~\bibnamefont
  {Wakeshima}},\ and\ \bibinfo {author} {\bibfnamefont {Y.}~\bibnamefont
  {Hinatsu}},\ }\bibfield  {title} {\bibinfo {title} {{Structural phase
  transition and magnetic properties of double perovskites Ba$_2$CaMO$_6$ ($M=$
  W, Re, Os)}},\ }\href
  {https://doi.org/https://doi.org/10.1016/j.jssc.2005.10.003} {\bibfield
  {journal} {\bibinfo  {journal} {J. Solid. State. Chem.}\ }\textbf {\bibinfo
  {volume} {179}},\ \bibinfo {pages} {605} (\bibinfo {year}
  {2006})}\BibitemShut {NoStop}%
\bibitem [{\citenamefont {Erickson}\ \emph {et~al.}(2007)\citenamefont
  {Erickson}, \citenamefont {Misra}, \citenamefont {Miller}, \citenamefont
  {Gupta}, \citenamefont {Schlesinger}, \citenamefont {Harrison}, \citenamefont
  {Kim},\ and\ \citenamefont {Fisher}}]{Erickson2007}%
  \BibitemOpen
  \bibfield  {author} {\bibinfo {author} {\bibfnamefont {A.~S.}\ \bibnamefont
  {Erickson}}, \bibinfo {author} {\bibfnamefont {S.}~\bibnamefont {Misra}},
  \bibinfo {author} {\bibfnamefont {G.~J.}\ \bibnamefont {Miller}}, \bibinfo
  {author} {\bibfnamefont {R.~R.}\ \bibnamefont {Gupta}}, \bibinfo {author}
  {\bibfnamefont {Z.}~\bibnamefont {Schlesinger}}, \bibinfo {author}
  {\bibfnamefont {W.~A.}\ \bibnamefont {Harrison}}, \bibinfo {author}
  {\bibfnamefont {J.~M.}\ \bibnamefont {Kim}},\ and\ \bibinfo {author}
  {\bibfnamefont {I.~R.}\ \bibnamefont {Fisher}},\ }\bibfield  {title}
  {\bibinfo {title} {{Ferromagnetism in the Mott Insulator
  ${\mathrm{Ba}}_{2}{\mathrm{NaOsO}}_{6}$}},\ }\href
  {https://doi.org/10.1103/PhysRevLett.99.016404} {\bibfield  {journal}
  {\bibinfo  {journal} {Phys. Rev. Lett.}\ }\textbf {\bibinfo {volume} {99}},\
  \bibinfo {pages} {016404} (\bibinfo {year} {2007})}\BibitemShut {NoStop}%
\bibitem [{\citenamefont {Steele}\ \emph {et~al.}(2011)\citenamefont {Steele},
  \citenamefont {Baker}, \citenamefont {Lancaster}, \citenamefont {Pratt},
  \citenamefont {Franke}, \citenamefont {Ghannadzadeh}, \citenamefont
  {Goddard}, \citenamefont {Hayes}, \citenamefont {Prabhakaran},\ and\
  \citenamefont {Blundell}}]{Steele2011}%
  \BibitemOpen
  \bibfield  {author} {\bibinfo {author} {\bibfnamefont {A.~J.}\ \bibnamefont
  {Steele}}, \bibinfo {author} {\bibfnamefont {P.~J.}\ \bibnamefont {Baker}},
  \bibinfo {author} {\bibfnamefont {T.}~\bibnamefont {Lancaster}}, \bibinfo
  {author} {\bibfnamefont {F.~L.}\ \bibnamefont {Pratt}}, \bibinfo {author}
  {\bibfnamefont {I.}~\bibnamefont {Franke}}, \bibinfo {author} {\bibfnamefont
  {S.}~\bibnamefont {Ghannadzadeh}}, \bibinfo {author} {\bibfnamefont {P.~A.}\
  \bibnamefont {Goddard}}, \bibinfo {author} {\bibfnamefont {W.}~\bibnamefont
  {Hayes}}, \bibinfo {author} {\bibfnamefont {D.}~\bibnamefont {Prabhakaran}},\
  and\ \bibinfo {author} {\bibfnamefont {S.~J.}\ \bibnamefont {Blundell}},\
  }\bibfield  {title} {\bibinfo {title} {{Low-moment magnetism in the double
  perovskites Ba${}_{2}$$M$OsO${}_{6}$ ($M=\text{Li},\text{Na}$)}},\ }\href
  {https://doi.org/10.1103/PhysRevB.84.144416} {\bibfield  {journal} {\bibinfo
  {journal} {Phys. Rev. B}\ }\textbf {\bibinfo {volume} {84}},\ \bibinfo
  {pages} {144416} (\bibinfo {year} {2011})}\BibitemShut {NoStop}%
\bibitem [{\citenamefont {Marjerrison}\ \emph {et~al.}(2016)\citenamefont
  {Marjerrison}, \citenamefont {Thompson}, \citenamefont {Sala}, \citenamefont
  {Maharaj}, \citenamefont {Kermarrec}, \citenamefont {Cai}, \citenamefont
  {Hallas}, \citenamefont {Wilson}, \citenamefont {Munsie}, \citenamefont
  {Granroth}, \citenamefont {Flacau}, \citenamefont {Greedan}, \citenamefont
  {Gaulin},\ and\ \citenamefont {Luke}}]{Marjerrison2016}%
  \BibitemOpen
  \bibfield  {author} {\bibinfo {author} {\bibfnamefont {C.~A.}\ \bibnamefont
  {Marjerrison}}, \bibinfo {author} {\bibfnamefont {C.~M.}\ \bibnamefont
  {Thompson}}, \bibinfo {author} {\bibfnamefont {G.}~\bibnamefont {Sala}},
  \bibinfo {author} {\bibfnamefont {D.~D.}\ \bibnamefont {Maharaj}}, \bibinfo
  {author} {\bibfnamefont {E.}~\bibnamefont {Kermarrec}}, \bibinfo {author}
  {\bibfnamefont {Y.}~\bibnamefont {Cai}}, \bibinfo {author} {\bibfnamefont
  {A.~M.}\ \bibnamefont {Hallas}}, \bibinfo {author} {\bibfnamefont {M.~N.}\
  \bibnamefont {Wilson}}, \bibinfo {author} {\bibfnamefont {T.~J.~S.}\
  \bibnamefont {Munsie}}, \bibinfo {author} {\bibfnamefont {G.~E.}\
  \bibnamefont {Granroth}}, \bibinfo {author} {\bibfnamefont {R.}~\bibnamefont
  {Flacau}}, \bibinfo {author} {\bibfnamefont {J.~E.}\ \bibnamefont {Greedan}},
  \bibinfo {author} {\bibfnamefont {B.~D.}\ \bibnamefont {Gaulin}},\ and\
  \bibinfo {author} {\bibfnamefont {G.~M.}\ \bibnamefont {Luke}},\ }\bibfield
  {title} {\bibinfo {title} {{Cubic Re$^{6+}$ ($5d^1$) Double Perovskites,
  Ba$_2$MgReO$_6$, Ba$_2$ZnReO$_6$, and Ba$_2$Y$_{2/3}$ReO$_6$: Magnetism, Heat
  Capacity, $\mu$SR, and Neutron Scattering Studies and Comparison with
  Theory}},\ }\href {https://doi.org/10.1021/acs.inorgchem.6b01933} {\bibfield
  {journal} {\bibinfo  {journal} {Inorg. Chem.}\ }\textbf {\bibinfo {volume}
  {55}},\ \bibinfo {pages} {10701} (\bibinfo {year} {2016})}\BibitemShut
  {NoStop}%
\bibitem [{\citenamefont {Lu}\ \emph {et~al.}(2017)\citenamefont {Lu},
  \citenamefont {Song}, \citenamefont {Liu}, \citenamefont {Reyes},
  \citenamefont {Kuhns}, \citenamefont {Lee}, \citenamefont {Fisher},\ and\
  \citenamefont {Mitrovi\'{c}}}]{Lu2017}%
  \BibitemOpen
  \bibfield  {author} {\bibinfo {author} {\bibfnamefont {L.}~\bibnamefont
  {Lu}}, \bibinfo {author} {\bibfnamefont {M.}~\bibnamefont {Song}}, \bibinfo
  {author} {\bibfnamefont {W.}~\bibnamefont {Liu}}, \bibinfo {author}
  {\bibfnamefont {A.~P.}\ \bibnamefont {Reyes}}, \bibinfo {author}
  {\bibfnamefont {P.}~\bibnamefont {Kuhns}}, \bibinfo {author} {\bibfnamefont
  {H.~O.}\ \bibnamefont {Lee}}, \bibinfo {author} {\bibfnamefont {I.~R.}\
  \bibnamefont {Fisher}},\ and\ \bibinfo {author} {\bibfnamefont {V.~F.}\
  \bibnamefont {Mitrovi\'{c}}},\ }\bibfield  {title} {\bibinfo {title}
  {{Magnetism and local symmetry breaking in a Mott insulator with strong spin
  orbit interactions}},\ }\href {https://doi.org/10.1038/ncomms14407}
  {\bibfield  {journal} {\bibinfo  {journal} {Nat. Commun.}\ }\textbf {\bibinfo
  {volume} {8}},\ \bibinfo {pages} {14407} (\bibinfo {year}
  {2017})}\BibitemShut {NoStop}%
\bibitem [{\citenamefont {Liu}\ \emph {et~al.}(2018)\citenamefont {Liu},
  \citenamefont {Cong}, \citenamefont {Garcia}, \citenamefont {Reyes},
  \citenamefont {Lee}, \citenamefont {Fisher},\ and\ \citenamefont
  {Mitrović}}]{Liu2018}%
  \BibitemOpen
  \bibfield  {author} {\bibinfo {author} {\bibfnamefont {W.}~\bibnamefont
  {Liu}}, \bibinfo {author} {\bibfnamefont {R.}~\bibnamefont {Cong}}, \bibinfo
  {author} {\bibfnamefont {E.}~\bibnamefont {Garcia}}, \bibinfo {author}
  {\bibfnamefont {A.}~\bibnamefont {Reyes}}, \bibinfo {author} {\bibfnamefont
  {H.}~\bibnamefont {Lee}}, \bibinfo {author} {\bibfnamefont {I.}~\bibnamefont
  {Fisher}},\ and\ \bibinfo {author} {\bibfnamefont {V.}~\bibnamefont
  {Mitrović}},\ }\bibfield  {title} {\bibinfo {title} {{Phase diagram of
  Ba$_2$NaOsO$_6$, a Mott insulator with strong spin orbit interactions}},\
  }\href {https://doi.org/https://doi.org/10.1016/j.physb.2017.08.062}
  {\bibfield  {journal} {\bibinfo  {journal} {Physica B: Condens. Matter}\
  }\textbf {\bibinfo {volume} {536}},\ \bibinfo {pages} {863} (\bibinfo {year}
  {2018})}\BibitemShut {NoStop}%
\bibitem [{\citenamefont {Willa}\ \emph {et~al.}(2019)\citenamefont {Willa},
  \citenamefont {Willa}, \citenamefont {Welp}, \citenamefont {Fisher},
  \citenamefont {Rydh}, \citenamefont {Kwok},\ and\ \citenamefont
  {Islam}}]{Willa2019}%
  \BibitemOpen
  \bibfield  {author} {\bibinfo {author} {\bibfnamefont {K.}~\bibnamefont
  {Willa}}, \bibinfo {author} {\bibfnamefont {R.}~\bibnamefont {Willa}},
  \bibinfo {author} {\bibfnamefont {U.}~\bibnamefont {Welp}}, \bibinfo {author}
  {\bibfnamefont {I.~R.}\ \bibnamefont {Fisher}}, \bibinfo {author}
  {\bibfnamefont {A.}~\bibnamefont {Rydh}}, \bibinfo {author} {\bibfnamefont
  {W.-K.}\ \bibnamefont {Kwok}},\ and\ \bibinfo {author} {\bibfnamefont
  {Z.}~\bibnamefont {Islam}},\ }\bibfield  {title} {\bibinfo {title} {{Phase
  transition preceding magnetic long-range order in the double perovskite
  ${\mathrm{Ba}}_{2}{\mathrm{NaOsO}}_{6}$}},\ }\href
  {https://doi.org/10.1103/PhysRevB.100.041108} {\bibfield  {journal} {\bibinfo
   {journal} {Phys. Rev. B}\ }\textbf {\bibinfo {volume} {100}},\ \bibinfo
  {pages} {041108(R)} (\bibinfo {year} {2019})}\BibitemShut {NoStop}%
\bibitem [{\citenamefont {Ishikawa}\ \emph {et~al.}(2019)\citenamefont
  {Ishikawa}, \citenamefont {Takayama}, \citenamefont {Kremer}, \citenamefont
  {Nuss}, \citenamefont {Dinnebier}, \citenamefont {Kitagawa}, \citenamefont
  {Ishii},\ and\ \citenamefont {Takagi}}]{Ishikawa2019}%
  \BibitemOpen
  \bibfield  {author} {\bibinfo {author} {\bibfnamefont {H.}~\bibnamefont
  {Ishikawa}}, \bibinfo {author} {\bibfnamefont {T.}~\bibnamefont {Takayama}},
  \bibinfo {author} {\bibfnamefont {R.~K.}\ \bibnamefont {Kremer}}, \bibinfo
  {author} {\bibfnamefont {J.}~\bibnamefont {Nuss}}, \bibinfo {author}
  {\bibfnamefont {R.}~\bibnamefont {Dinnebier}}, \bibinfo {author}
  {\bibfnamefont {K.}~\bibnamefont {Kitagawa}}, \bibinfo {author}
  {\bibfnamefont {K.}~\bibnamefont {Ishii}},\ and\ \bibinfo {author}
  {\bibfnamefont {H.}~\bibnamefont {Takagi}},\ }\bibfield  {title} {\bibinfo
  {title} {{Ordering of hidden multipoles in spin-orbit entangled $5{d}^{1}$ Ta
  chlorides}},\ }\href {https://doi.org/10.1103/PhysRevB.100.045142} {\bibfield
   {journal} {\bibinfo  {journal} {Phys. Rev. B}\ }\textbf {\bibinfo {volume}
  {100}},\ \bibinfo {pages} {045142} (\bibinfo {year} {2019})}\BibitemShut
  {NoStop}%
\bibitem [{\citenamefont {Hirai}\ and\ \citenamefont
  {Hiroi}(2019)}]{Hirai2019}%
  \BibitemOpen
  \bibfield  {author} {\bibinfo {author} {\bibfnamefont {D.}~\bibnamefont
  {Hirai}}\ and\ \bibinfo {author} {\bibfnamefont {Z.}~\bibnamefont {Hiroi}},\
  }\bibfield  {title} {\bibinfo {title} {{Successive Symmetry Breaking in a
  $J_\text{eff}$ = 3/2 Quartet in the Spin-Orbit Coupled Insulator
  Ba$_2$MgReO$_6$}},\ }\href {https://doi.org/10.7566/JPSJ.88.064712}
  {\bibfield  {journal} {\bibinfo  {journal} {J. Phys. Soc. Jpn.}\ }\textbf
  {\bibinfo {volume} {88}},\ \bibinfo {pages} {064712} (\bibinfo {year}
  {2019})}\BibitemShut {NoStop}%
\bibitem [{\citenamefont {Hirai}\ \emph {et~al.}(2020)\citenamefont {Hirai},
  \citenamefont {Sagayama}, \citenamefont {Gao}, \citenamefont {Ohsumi},
  \citenamefont {Chen}, \citenamefont {Arima},\ and\ \citenamefont
  {Hiroi}}]{Hirai2020}%
  \BibitemOpen
  \bibfield  {author} {\bibinfo {author} {\bibfnamefont {D.}~\bibnamefont
  {Hirai}}, \bibinfo {author} {\bibfnamefont {H.}~\bibnamefont {Sagayama}},
  \bibinfo {author} {\bibfnamefont {S.}~\bibnamefont {Gao}}, \bibinfo {author}
  {\bibfnamefont {H.}~\bibnamefont {Ohsumi}}, \bibinfo {author} {\bibfnamefont
  {G.}~\bibnamefont {Chen}}, \bibinfo {author} {\bibfnamefont {T.-h.}\
  \bibnamefont {Arima}},\ and\ \bibinfo {author} {\bibfnamefont
  {Z.}~\bibnamefont {Hiroi}},\ }\bibfield  {title} {\bibinfo {title}
  {{Detection of multipolar orders in the spin-orbit-coupled $5d$ Mott
  insulator $\mathrm{B}{\mathrm{a}}_{2}\mathrm{MgRe}{\mathrm{O}}_{6}$}},\
  }\href {https://doi.org/10.1103/PhysRevResearch.2.022063} {\bibfield
  {journal} {\bibinfo  {journal} {Phys. Rev. Research}\ }\textbf {\bibinfo
  {volume} {2}},\ \bibinfo {pages} {022063(R)} (\bibinfo {year}
  {2020})}\BibitemShut {NoStop}%
\bibitem [{\citenamefont {Hirai}\ and\ \citenamefont
  {Hiroi}(2021)}]{Hirai2021}%
  \BibitemOpen
  \bibfield  {author} {\bibinfo {author} {\bibfnamefont {D.}~\bibnamefont
  {Hirai}}\ and\ \bibinfo {author} {\bibfnamefont {Z.}~\bibnamefont {Hiroi}},\
  }\bibfield  {title} {\bibinfo {title} {{Possible quadrupole order in
  tetragonal Ba$_2$CdReO$_6$ and chemical trend in the ground states of 5d$^1$
  double perovskites}},\ }\href {https://doi.org/10.1088/1361-648x/abda79}
  {\bibfield  {journal} {\bibinfo  {journal} {J. Phys.: Condens. Matter}\
  }\textbf {\bibinfo {volume} {33}},\ \bibinfo {pages} {135603} (\bibinfo
  {year} {2021})}\BibitemShut {NoStop}%
\bibitem [{\citenamefont {Ishikawa}\ \emph
  {et~al.}(2021{\natexlab{a}})\citenamefont {Ishikawa}, \citenamefont {Yajima},
  \citenamefont {Matsuo},\ and\ \citenamefont {Kindo}}]{Ishikawa2021}%
  \BibitemOpen
  \bibfield  {author} {\bibinfo {author} {\bibfnamefont {H.}~\bibnamefont
  {Ishikawa}}, \bibinfo {author} {\bibfnamefont {T.}~\bibnamefont {Yajima}},
  \bibinfo {author} {\bibfnamefont {A.}~\bibnamefont {Matsuo}},\ and\ \bibinfo
  {author} {\bibfnamefont {K.}~\bibnamefont {Kindo}},\ }\bibfield  {title}
  {\bibinfo {title} {{Ligand dependent magnetism of the $J_\text{eff} = 3/2$
  Mott insulator Cs$_2MX_6$ ($M =$ Ta, Nb, $X =$ Br, Cl)}},\ }\href
  {https://doi.org/10.1088/1361-648x/abd7b5} {\bibfield  {journal} {\bibinfo
  {journal} {J. Phys.: Condens. Matter}\ }\textbf {\bibinfo {volume} {33}},\
  \bibinfo {pages} {125802} (\bibinfo {year} {2021}{\natexlab{a}})}\BibitemShut
  {NoStop}%
\bibitem [{\citenamefont {Ishikawa}\ \emph
  {et~al.}(2021{\natexlab{b}})\citenamefont {Ishikawa}, \citenamefont {Hirai},
  \citenamefont {Ikeda}, \citenamefont {Gen}, \citenamefont {Yajima},
  \citenamefont {Matsuo}, \citenamefont {Matsuda}, \citenamefont {Hiroi},\ and\
  \citenamefont {Kindo}}]{Ishikawa2021b}%
  \BibitemOpen
  \bibfield  {author} {\bibinfo {author} {\bibfnamefont {H.}~\bibnamefont
  {Ishikawa}}, \bibinfo {author} {\bibfnamefont {D.}~\bibnamefont {Hirai}},
  \bibinfo {author} {\bibfnamefont {A.}~\bibnamefont {Ikeda}}, \bibinfo
  {author} {\bibfnamefont {M.}~\bibnamefont {Gen}}, \bibinfo {author}
  {\bibfnamefont {T.}~\bibnamefont {Yajima}}, \bibinfo {author} {\bibfnamefont
  {A.}~\bibnamefont {Matsuo}}, \bibinfo {author} {\bibfnamefont {Y.~H.}\
  \bibnamefont {Matsuda}}, \bibinfo {author} {\bibfnamefont {Z.}~\bibnamefont
  {Hiroi}},\ and\ \bibinfo {author} {\bibfnamefont {K.}~\bibnamefont {Kindo}},\
  }\bibfield  {title} {\bibinfo {title} {{Phase transition in the $5{d}^{1}$
  double perovskite ${\mathrm{Ba}}_{2}{\mathrm{CaReO}}_{6}$ induced by high
  magnetic field}},\ }\href {https://doi.org/10.1103/PhysRevB.104.174422}
  {\bibfield  {journal} {\bibinfo  {journal} {Phys. Rev. B}\ }\textbf {\bibinfo
  {volume} {104}},\ \bibinfo {pages} {174422} (\bibinfo {year}
  {2021}{\natexlab{b}})}\BibitemShut {NoStop}%
\bibitem [{\citenamefont {Arima}\ \emph {et~al.}(2022)\citenamefont {Arima},
  \citenamefont {Oshita}, \citenamefont {Hirai}, \citenamefont {Hiroi},\ and\
  \citenamefont {Matsubayashi}}]{Arima2022}%
  \BibitemOpen
  \bibfield  {author} {\bibinfo {author} {\bibfnamefont {H.}~\bibnamefont
  {Arima}}, \bibinfo {author} {\bibfnamefont {Y.}~\bibnamefont {Oshita}},
  \bibinfo {author} {\bibfnamefont {D.}~\bibnamefont {Hirai}}, \bibinfo
  {author} {\bibfnamefont {Z.}~\bibnamefont {Hiroi}},\ and\ \bibinfo {author}
  {\bibfnamefont {K.}~\bibnamefont {Matsubayashi}},\ }\bibfield  {title}
  {\bibinfo {title} {{Interplay between Quadrupolar and Magnetic Interactions
  in $5d^1$ Double Perovskite Ba$_2$MgReO$_6$ under Pressure}},\ }\href
  {https://doi.org/10.7566/JPSJ.91.013702} {\bibfield  {journal} {\bibinfo
  {journal} {J. Phys. Soc. Jpn.}\ }\textbf {\bibinfo {volume} {91}},\ \bibinfo
  {pages} {013702} (\bibinfo {year} {2022})}\BibitemShut {NoStop}%
\bibitem [{\citenamefont {da~Cruz Pinha~Barbosa}\ \emph
  {et~al.}(2022)\citenamefont {da~Cruz Pinha~Barbosa}, \citenamefont {Xiong},
  \citenamefont {Tran}, \citenamefont {McGuire}, \citenamefont {Yan},
  \citenamefont {Warren}, \citenamefont {Aguilar}, \citenamefont {Zhang},
  \citenamefont {Randeria}, \citenamefont {Trivedi}, \citenamefont {Haskel},\
  and\ \citenamefont {Woodward}}]{Woodward2022}%
  \BibitemOpen
  \bibfield  {author} {\bibinfo {author} {\bibfnamefont {V.}~\bibnamefont
  {da~Cruz Pinha~Barbosa}}, \bibinfo {author} {\bibfnamefont {J.}~\bibnamefont
  {Xiong}}, \bibinfo {author} {\bibfnamefont {P.~M.}\ \bibnamefont {Tran}},
  \bibinfo {author} {\bibfnamefont {M.~A.}\ \bibnamefont {McGuire}}, \bibinfo
  {author} {\bibfnamefont {J.}~\bibnamefont {Yan}}, \bibinfo {author}
  {\bibfnamefont {M.~T.}\ \bibnamefont {Warren}}, \bibinfo {author}
  {\bibfnamefont {R.~V.}\ \bibnamefont {Aguilar}}, \bibinfo {author}
  {\bibfnamefont {W.}~\bibnamefont {Zhang}}, \bibinfo {author} {\bibfnamefont
  {M.}~\bibnamefont {Randeria}}, \bibinfo {author} {\bibfnamefont
  {N.}~\bibnamefont {Trivedi}}, \bibinfo {author} {\bibfnamefont
  {D.}~\bibnamefont {Haskel}},\ and\ \bibinfo {author} {\bibfnamefont {P.~M.}\
  \bibnamefont {Woodward}},\ }\bibfield  {title} {\bibinfo {title} {{The Impact
  of Structural Distortions on the Magnetism of Double Perovskites Containing
  $5d^1$ Transition-Metal Ions}},\ }\href
  {https://doi.org/10.1021/acs.chemmater.1c03456} {\bibfield  {journal}
  {\bibinfo  {journal} {Chem. Mater.}\ }\textbf {\bibinfo {volume} {34}},\
  \bibinfo {pages} {1098} (\bibinfo {year} {2022})}\BibitemShut {NoStop}%
\bibitem [{\citenamefont {Cong}(2022)}]{Cong2022}%
  \BibitemOpen
  \bibfield  {author} {\bibinfo {author} {\bibfnamefont {R.}~\bibnamefont
  {Cong}},\ }\emph {\bibinfo {title} {Magnetic and Structural Properties of 5d
  Osmate Double Perovskites Probed by Nuclear Magnetic Resonance}},\ \href
  {https://doi.org/10.26300/7tx8-gq83} {Ph.D. thesis},\ \bibinfo  {school}
  {Brown University} (\bibinfo {year} {2022})\BibitemShut {NoStop}%
\bibitem [{\citenamefont {Mansouri~Tehrani}\ \emph {et~al.}(2023)\citenamefont
  {Mansouri~Tehrani}, \citenamefont {Soh}, \citenamefont {P\'asztorov\'a},
  \citenamefont {Merkel}, \citenamefont {\ifmmode \check{Z}\else
  \v{Z}\fi{}ivkovi\ifmmode~\acute{c}\else \'{c}\fi{}}, \citenamefont
  {R\o{}nnow},\ and\ \citenamefont {Spaldin}}]{Tehrani2023}%
  \BibitemOpen
  \bibfield  {author} {\bibinfo {author} {\bibfnamefont {A.}~\bibnamefont
  {Mansouri~Tehrani}}, \bibinfo {author} {\bibfnamefont {J.-R.}\ \bibnamefont
  {Soh}}, \bibinfo {author} {\bibfnamefont {J.}~\bibnamefont {P\'asztorov\'a}},
  \bibinfo {author} {\bibfnamefont {M.~E.}\ \bibnamefont {Merkel}}, \bibinfo
  {author} {\bibfnamefont {I.}~\bibnamefont {\ifmmode \check{Z}\else
  \v{Z}\fi{}ivkovi\ifmmode~\acute{c}\else \'{c}\fi{}}}, \bibinfo {author}
  {\bibfnamefont {H.~M.}\ \bibnamefont {R\o{}nnow}},\ and\ \bibinfo {author}
  {\bibfnamefont {N.~A.}\ \bibnamefont {Spaldin}},\ }\bibfield  {title}
  {\bibinfo {title} {{Charge multipole correlations and order in
  ${\mathrm{Cs}}_{2}\mathrm{Ta}{\mathrm{Cl}}_{6}$}},\ }\href
  {https://doi.org/10.1103/PhysRevResearch.5.L012010} {\bibfield  {journal}
  {\bibinfo  {journal} {Phys. Rev. Res.}\ }\textbf {\bibinfo {volume} {5}},\
  \bibinfo {pages} {L012010} (\bibinfo {year} {2023})}\BibitemShut {NoStop}%
\bibitem [{\citenamefont {Cussen}\ \emph {et~al.}(2006)\citenamefont {Cussen},
  \citenamefont {Lynham},\ and\ \citenamefont {Rogers}}]{Cussen2006}%
  \BibitemOpen
  \bibfield  {author} {\bibinfo {author} {\bibfnamefont {E.~J.}\ \bibnamefont
  {Cussen}}, \bibinfo {author} {\bibfnamefont {D.~R.}\ \bibnamefont {Lynham}},\
  and\ \bibinfo {author} {\bibfnamefont {J.}~\bibnamefont {Rogers}},\
  }\bibfield  {title} {\bibinfo {title} {{Magnetic Order Arising from
  Structural Distortion: Structure and Magnetic Properties of
  Ba$_2$LnMoO$_6$}},\ }\href {https://doi.org/10.1021/cm0602388} {\bibfield
  {journal} {\bibinfo  {journal} {Chem. Mater.}\ }\textbf {\bibinfo {volume}
  {18}},\ \bibinfo {pages} {2855} (\bibinfo {year} {2006})}\BibitemShut
  {NoStop}%
\bibitem [{\citenamefont {Aharen}\ \emph {et~al.}(2010)\citenamefont {Aharen},
  \citenamefont {Greedan}, \citenamefont {Bridges}, \citenamefont {Aczel},
  \citenamefont {Rodriguez}, \citenamefont {MacDougall}, \citenamefont {Luke},
  \citenamefont {Imai}, \citenamefont {Michaelis}, \citenamefont {Kroeker},
  \citenamefont {Zhou}, \citenamefont {Wiebe},\ and\ \citenamefont
  {Cranswick}}]{Aharen2010}%
  \BibitemOpen
  \bibfield  {author} {\bibinfo {author} {\bibfnamefont {T.}~\bibnamefont
  {Aharen}}, \bibinfo {author} {\bibfnamefont {J.~E.}\ \bibnamefont {Greedan}},
  \bibinfo {author} {\bibfnamefont {C.~A.}\ \bibnamefont {Bridges}}, \bibinfo
  {author} {\bibfnamefont {A.~A.}\ \bibnamefont {Aczel}}, \bibinfo {author}
  {\bibfnamefont {J.}~\bibnamefont {Rodriguez}}, \bibinfo {author}
  {\bibfnamefont {G.}~\bibnamefont {MacDougall}}, \bibinfo {author}
  {\bibfnamefont {G.~M.}\ \bibnamefont {Luke}}, \bibinfo {author}
  {\bibfnamefont {T.}~\bibnamefont {Imai}}, \bibinfo {author} {\bibfnamefont
  {V.~K.}\ \bibnamefont {Michaelis}}, \bibinfo {author} {\bibfnamefont
  {S.}~\bibnamefont {Kroeker}}, \bibinfo {author} {\bibfnamefont
  {H.}~\bibnamefont {Zhou}}, \bibinfo {author} {\bibfnamefont {C.~R.}\
  \bibnamefont {Wiebe}},\ and\ \bibinfo {author} {\bibfnamefont {L.~M.~D.}\
  \bibnamefont {Cranswick}},\ }\bibfield  {title} {\bibinfo {title} {{Magnetic
  properties of the geometrically frustrated $S=\frac{1}{2}$ antiferromagnets,
  ${\text{La}}_{2}{\text{LiMoO}}_{6}$ and ${\text{Ba}}_{2}{\text{YMoO}}_{6}$,
  with the B-site ordered double perovskite structure: Evidence for a
  collective spin-singlet ground state}},\ }\href
  {https://doi.org/10.1103/PhysRevB.81.224409} {\bibfield  {journal} {\bibinfo
  {journal} {Phys. Rev. B}\ }\textbf {\bibinfo {volume} {81}},\ \bibinfo
  {pages} {224409} (\bibinfo {year} {2010})}\BibitemShut {NoStop}%
\bibitem [{\citenamefont {de~Vries}\ \emph {et~al.}(2010)\citenamefont
  {de~Vries}, \citenamefont {Mclaughlin},\ and\ \citenamefont
  {Bos}}]{deVries2010}%
  \BibitemOpen
  \bibfield  {author} {\bibinfo {author} {\bibfnamefont {M.~A.}\ \bibnamefont
  {de~Vries}}, \bibinfo {author} {\bibfnamefont {A.~C.}\ \bibnamefont
  {Mclaughlin}},\ and\ \bibinfo {author} {\bibfnamefont {J.~W.~G.}\
  \bibnamefont {Bos}},\ }\bibfield  {title} {\bibinfo {title} {{Valence Bond
  Glass on an fcc Lattice in the Double Perovskite
  ${\mathrm{Ba}}_{2}{\mathrm{YMoO}}_{6}$}},\ }\href
  {https://doi.org/10.1103/PhysRevLett.104.177202} {\bibfield  {journal}
  {\bibinfo  {journal} {Phys. Rev. Lett.}\ }\textbf {\bibinfo {volume} {104}},\
  \bibinfo {pages} {177202} (\bibinfo {year} {2010})}\BibitemShut {NoStop}%
\bibitem [{\citenamefont {Coomer}\ and\ \citenamefont
  {Cussen}(2013)}]{Coomer2013}%
  \BibitemOpen
  \bibfield  {author} {\bibinfo {author} {\bibfnamefont {F.~C.}\ \bibnamefont
  {Coomer}}\ and\ \bibinfo {author} {\bibfnamefont {E.~J.}\ \bibnamefont
  {Cussen}},\ }\bibfield  {title} {\bibinfo {title} {{Structural and magnetic
  properties of Ba$_2$LuMoO$_6$: a valence bond glass}},\ }\href
  {https://doi.org/10.1088/0953-8984/25/8/082202} {\bibfield  {journal}
  {\bibinfo  {journal} {J. Phys.: Condens. Matter}\ }\textbf {\bibinfo {volume}
  {25}},\ \bibinfo {pages} {082202} (\bibinfo {year} {2013})}\BibitemShut
  {NoStop}%
\bibitem [{\citenamefont {Lee}\ \emph {et~al.}(2021)\citenamefont {Lee},
  \citenamefont {Lee}, \citenamefont {Guohua}, \citenamefont {Ma},
  \citenamefont {Zhou}, \citenamefont {Lee}, \citenamefont {Choi},\ and\
  \citenamefont {Choi}}]{Lee2021}%
  \BibitemOpen
  \bibfield  {author} {\bibinfo {author} {\bibfnamefont {S.}~\bibnamefont
  {Lee}}, \bibinfo {author} {\bibfnamefont {W.}~\bibnamefont {Lee}}, \bibinfo
  {author} {\bibfnamefont {W.}~\bibnamefont {Guohua}}, \bibinfo {author}
  {\bibfnamefont {J.}~\bibnamefont {Ma}}, \bibinfo {author} {\bibfnamefont
  {H.}~\bibnamefont {Zhou}}, \bibinfo {author} {\bibfnamefont {M.}~\bibnamefont
  {Lee}}, \bibinfo {author} {\bibfnamefont {E.~S.}\ \bibnamefont {Choi}},\ and\
  \bibinfo {author} {\bibfnamefont {K.-Y.}\ \bibnamefont {Choi}},\ }\bibfield
  {title} {\bibinfo {title} {{Experimental evidence for a valence-bond glass in
  the $5{d}^{1}$ double perovskite ${\mathrm{Ba}}_{2}{\mathrm{YWO}}_{6}$}},\
  }\href {https://doi.org/10.1103/PhysRevB.103.224430} {\bibfield  {journal}
  {\bibinfo  {journal} {Phys. Rev. B}\ }\textbf {\bibinfo {volume} {103}},\
  \bibinfo {pages} {224430} (\bibinfo {year} {2021})}\BibitemShut {NoStop}%
\bibitem [{\citenamefont {Mustonen}\ \emph {et~al.}(2022)\citenamefont
  {Mustonen}, \citenamefont {Mutch}, \citenamefont {Walker}, \citenamefont
  {Baker}, \citenamefont {Coomer}, \citenamefont {Perry}, \citenamefont
  {Pughe}, \citenamefont {Stenning}, \citenamefont {Liu}, \citenamefont
  {Dutton},\ and\ \citenamefont {Cussen}}]{Mustonen2022}%
  \BibitemOpen
  \bibfield  {author} {\bibinfo {author} {\bibfnamefont {O.~H.}\ \bibnamefont
  {Mustonen}}, \bibinfo {author} {\bibfnamefont {H.~N.}\ \bibnamefont {Mutch}},
  \bibinfo {author} {\bibfnamefont {H.~C.}\ \bibnamefont {Walker}}, \bibinfo
  {author} {\bibfnamefont {P.~J.}\ \bibnamefont {Baker}}, \bibinfo {author}
  {\bibfnamefont {F.~C.}\ \bibnamefont {Coomer}}, \bibinfo {author}
  {\bibfnamefont {R.~S.}\ \bibnamefont {Perry}}, \bibinfo {author}
  {\bibfnamefont {C.}~\bibnamefont {Pughe}}, \bibinfo {author} {\bibfnamefont
  {G.~B.~G.}\ \bibnamefont {Stenning}}, \bibinfo {author} {\bibfnamefont
  {C.}~\bibnamefont {Liu}}, \bibinfo {author} {\bibfnamefont {S.~E.}\
  \bibnamefont {Dutton}},\ and\ \bibinfo {author} {\bibfnamefont {E.~J.}\
  \bibnamefont {Cussen}},\ }\bibfield  {title} {\bibinfo {title} {{Valence bond
  glass state in the 4$d^1$ fcc antiferromagnet Ba$_2$LuMoO$_6$}},\ }\href
  {https://doi.org/10.1038/s41535-022-00480-4} {\bibfield  {journal} {\bibinfo
  {journal} {npj Quantum Mater.}\ }\textbf {\bibinfo {volume} {7}},\ \bibinfo
  {pages} {74} (\bibinfo {year} {2022})}\BibitemShut {NoStop}%
\bibitem [{\citenamefont {Chen}\ \emph {et~al.}(2010)\citenamefont {Chen},
  \citenamefont {Pereira},\ and\ \citenamefont {Balents}}]{Chen2010}%
  \BibitemOpen
  \bibfield  {author} {\bibinfo {author} {\bibfnamefont {G.}~\bibnamefont
  {Chen}}, \bibinfo {author} {\bibfnamefont {R.}~\bibnamefont {Pereira}},\ and\
  \bibinfo {author} {\bibfnamefont {L.}~\bibnamefont {Balents}},\ }\bibfield
  {title} {\bibinfo {title} {Exotic phases induced by strong spin-orbit
  coupling in ordered double perovskites},\ }\href
  {https://doi.org/10.1103/PhysRevB.82.174440} {\bibfield  {journal} {\bibinfo
  {journal} {Phys. Rev. B}\ }\textbf {\bibinfo {volume} {82}},\ \bibinfo
  {pages} {174440} (\bibinfo {year} {2010})}\BibitemShut {NoStop}%
\bibitem [{\citenamefont {Romh\'anyi}\ \emph {et~al.}(2017)\citenamefont
  {Romh\'anyi}, \citenamefont {Balents},\ and\ \citenamefont
  {Jackeli}}]{Romhanyi2017}%
  \BibitemOpen
  \bibfield  {author} {\bibinfo {author} {\bibfnamefont {J.}~\bibnamefont
  {Romh\'anyi}}, \bibinfo {author} {\bibfnamefont {L.}~\bibnamefont
  {Balents}},\ and\ \bibinfo {author} {\bibfnamefont {G.}~\bibnamefont
  {Jackeli}},\ }\bibfield  {title} {\bibinfo {title} {{Spin-Orbit Dimers and
  Noncollinear Phases in ${d}^{1}$ Cubic Double Perovskites}},\ }\href
  {https://doi.org/10.1103/PhysRevLett.118.217202} {\bibfield  {journal}
  {\bibinfo  {journal} {Phys. Rev. Lett.}\ }\textbf {\bibinfo {volume} {118}},\
  \bibinfo {pages} {217202} (\bibinfo {year} {2017})}\BibitemShut {NoStop}%
\bibitem [{\citenamefont {Svoboda}\ \emph {et~al.}(2021)\citenamefont
  {Svoboda}, \citenamefont {Zhang}, \citenamefont {Randeria},\ and\
  \citenamefont {Trivedi}}]{Svoboda2021}%
  \BibitemOpen
  \bibfield  {author} {\bibinfo {author} {\bibfnamefont {C.}~\bibnamefont
  {Svoboda}}, \bibinfo {author} {\bibfnamefont {W.}~\bibnamefont {Zhang}},
  \bibinfo {author} {\bibfnamefont {M.}~\bibnamefont {Randeria}},\ and\
  \bibinfo {author} {\bibfnamefont {N.}~\bibnamefont {Trivedi}},\ }\bibfield
  {title} {\bibinfo {title} {Orbital order drives magnetic order in $5{d}^{1}$
  and $5{d}^{2}$ double perovskite mott insulators},\ }\href
  {https://doi.org/10.1103/PhysRevB.104.024437} {\bibfield  {journal} {\bibinfo
   {journal} {Phys. Rev. B}\ }\textbf {\bibinfo {volume} {104}},\ \bibinfo
  {pages} {024437} (\bibinfo {year} {2021})}\BibitemShut {NoStop}%
\bibitem [{\citenamefont {Iwahara}\ \emph {et~al.}(2018)\citenamefont
  {Iwahara}, \citenamefont {Vieru},\ and\ \citenamefont
  {Chibotaru}}]{Iwahara2018}%
  \BibitemOpen
  \bibfield  {author} {\bibinfo {author} {\bibfnamefont {N.}~\bibnamefont
  {Iwahara}}, \bibinfo {author} {\bibfnamefont {V.}~\bibnamefont {Vieru}},\
  and\ \bibinfo {author} {\bibfnamefont {L.~F.}\ \bibnamefont {Chibotaru}},\
  }\bibfield  {title} {\bibinfo {title} {Spin-orbital-lattice entangled states
  in cubic ${d}^{1}$ double perovskites},\ }\href
  {https://doi.org/10.1103/PhysRevB.98.075138} {\bibfield  {journal} {\bibinfo
  {journal} {Phys. Rev. B}\ }\textbf {\bibinfo {volume} {98}},\ \bibinfo
  {pages} {075138} (\bibinfo {year} {2018})}\BibitemShut {NoStop}%
\bibitem [{\citenamefont {Xu}\ \emph {et~al.}(2016)\citenamefont {Xu},
  \citenamefont {Bogdanov}, \citenamefont {Princep}, \citenamefont {Fulde},
  \citenamefont {van~den Brink},\ and\ \citenamefont {Hozoi}}]{Xu2016}%
  \BibitemOpen
  \bibfield  {author} {\bibinfo {author} {\bibfnamefont {L.}~\bibnamefont
  {Xu}}, \bibinfo {author} {\bibfnamefont {N.~A.}\ \bibnamefont {Bogdanov}},
  \bibinfo {author} {\bibfnamefont {A.}~\bibnamefont {Princep}}, \bibinfo
  {author} {\bibfnamefont {P.}~\bibnamefont {Fulde}}, \bibinfo {author}
  {\bibfnamefont {J.}~\bibnamefont {van~den Brink}},\ and\ \bibinfo {author}
  {\bibfnamefont {L.}~\bibnamefont {Hozoi}},\ }\bibfield  {title} {\bibinfo
  {title} {Covalency and vibronic couplings make a nonmagnetic $j=3/2$ ion
  magnetic},\ }\href {https://doi.org/10.1038/npjquantmats.2016.29} {\bibfield
  {journal} {\bibinfo  {journal} {npj Quantum Mater.}\ }\textbf {\bibinfo
  {volume} {1}},\ \bibinfo {pages} {16029} (\bibinfo {year}
  {2016})}\BibitemShut {NoStop}%
\bibitem [{\citenamefont {Sugano}\ \emph {et~al.}(1970)\citenamefont {Sugano},
  \citenamefont {Tanabe},\ and\ \citenamefont {Kamimura}}]{Sugano1970}%
  \BibitemOpen
  \bibfield  {author} {\bibinfo {author} {\bibfnamefont {S.}~\bibnamefont
  {Sugano}}, \bibinfo {author} {\bibfnamefont {Y.}~\bibnamefont {Tanabe}},\
  and\ \bibinfo {author} {\bibfnamefont {H.}~\bibnamefont {Kamimura}},\
  }\href@noop {} {\emph {\bibinfo {title} {{Multiplets of Transition-Metal Ions
  in Crystals}}}}\ (\bibinfo  {publisher} {Academic Press},\ \bibinfo {address}
  {New York},\ \bibinfo {year} {1970})\BibitemShut {NoStop}%
\bibitem [{\citenamefont {Bersuker}\ and\ \citenamefont
  {Polinger}(1989)}]{Bersuker1989}%
  \BibitemOpen
  \bibfield  {author} {\bibinfo {author} {\bibfnamefont {I.~B.}\ \bibnamefont
  {Bersuker}}\ and\ \bibinfo {author} {\bibfnamefont {V.~Z.}\ \bibnamefont
  {Polinger}},\ }\href {https://doi.org/10.1007/978-3-642-83479-0} {\emph
  {\bibinfo {title} {Vibronic Interactions in Molecules and Crystals}}}\
  (\bibinfo  {publisher} {Springer-Verlag},\ \bibinfo {address} {Berlin and
  Heidelberg},\ \bibinfo {year} {1989})\BibitemShut {NoStop}%
\bibitem [{\citenamefont {Mosca}\ \emph {et~al.}(2023)\citenamefont {Mosca},
  \citenamefont {Schnait}, \citenamefont {Celiberti}, \citenamefont
  {Aichhorn},\ and\ \citenamefont {Franchini}}]{Mosca2023}%
  \BibitemOpen
  \bibfield  {author} {\bibinfo {author} {\bibfnamefont {D.~F.}\ \bibnamefont
  {Mosca}}, \bibinfo {author} {\bibfnamefont {H.}~\bibnamefont {Schnait}},
  \bibinfo {author} {\bibfnamefont {L.}~\bibnamefont {Celiberti}}, \bibinfo
  {author} {\bibfnamefont {M.}~\bibnamefont {Aichhorn}},\ and\ \bibinfo
  {author} {\bibfnamefont {C.}~\bibnamefont {Franchini}},\ }\href@noop {}
  {\bibinfo {title} {{The Mott transition in the 5d$^1$ compound
  Ba$_2$NaOsO$_6:$ a DFT+DMFT study with PAW non-collinear projectors}}}
  (\bibinfo {year} {2023}),\ \bibinfo {note} {arXiv:2303.16560}\BibitemShut
  {NoStop}%
\bibitem [{\citenamefont {Natori}\ \emph {et~al.}(2016)\citenamefont {Natori},
  \citenamefont {Andrade}, \citenamefont {Miranda},\ and\ \citenamefont
  {Pereira}}]{Natori2016}%
  \BibitemOpen
  \bibfield  {author} {\bibinfo {author} {\bibfnamefont {W.~M.~H.}\
  \bibnamefont {Natori}}, \bibinfo {author} {\bibfnamefont {E.~C.}\
  \bibnamefont {Andrade}}, \bibinfo {author} {\bibfnamefont {E.}~\bibnamefont
  {Miranda}},\ and\ \bibinfo {author} {\bibfnamefont {R.~G.}\ \bibnamefont
  {Pereira}},\ }\bibfield  {title} {\bibinfo {title} {Chiral spin-orbital
  liquids with nodal lines},\ }\href
  {https://doi.org/10.1103/PhysRevLett.117.017204} {\bibfield  {journal}
  {\bibinfo  {journal} {Phys. Rev. Lett.}\ }\textbf {\bibinfo {volume} {117}},\
  \bibinfo {pages} {017204} (\bibinfo {year} {2016})}\BibitemShut {NoStop}%
\bibitem [{\citenamefont {Ham}(1987)}]{Ham1987}%
  \BibitemOpen
  \bibfield  {author} {\bibinfo {author} {\bibfnamefont {F.~S.}\ \bibnamefont
  {Ham}},\ }\bibfield  {title} {\bibinfo {title} {{Berry's geometrical phase
  and the sequence of states in the Jahn-Teller effect}},\ }\href
  {https://doi.org/10.1103/PhysRevLett.58.725} {\bibfield  {journal} {\bibinfo
  {journal} {Phys. Rev. Lett.}\ }\textbf {\bibinfo {volume} {58}},\ \bibinfo
  {pages} {725} (\bibinfo {year} {1987})}\BibitemShut {NoStop}%
\bibitem [{\citenamefont {Tsunetsugu}\ \emph {et~al.}(2021)\citenamefont
  {Tsunetsugu}, \citenamefont {Ishitobi},\ and\ \citenamefont
  {Hattori}}]{Tsunetsugu2021}%
  \BibitemOpen
  \bibfield  {author} {\bibinfo {author} {\bibfnamefont {H.}~\bibnamefont
  {Tsunetsugu}}, \bibinfo {author} {\bibfnamefont {T.}~\bibnamefont
  {Ishitobi}},\ and\ \bibinfo {author} {\bibfnamefont {K.}~\bibnamefont
  {Hattori}},\ }\bibfield  {title} {\bibinfo {title} {Quadrupole orders on the
  fcc lattice},\ }\href {https://doi.org/10.7566/JPSJ.90.043701} {\bibfield
  {journal} {\bibinfo  {journal} {J. Phys. Soc. Jpn.}\ }\textbf {\bibinfo
  {volume} {90}},\ \bibinfo {pages} {043701} (\bibinfo {year}
  {2021})}\BibitemShut {NoStop}%
\bibitem [{\citenamefont {Ishihara}\ and\ \citenamefont
  {Maekawa}(2002)}]{Ishihara2002}%
  \BibitemOpen
  \bibfield  {author} {\bibinfo {author} {\bibfnamefont {S.}~\bibnamefont
  {Ishihara}}\ and\ \bibinfo {author} {\bibfnamefont {S.}~\bibnamefont
  {Maekawa}},\ }\bibfield  {title} {\bibinfo {title} {Resonant x-ray scattering
  in manganites: study of the orbital degree of freedom},\ }\href
  {https://doi.org/10.1088/0034-4885/65/4/203} {\bibfield  {journal} {\bibinfo
  {journal} {Rep. Prog. Phys.}\ }\textbf {\bibinfo {volume} {65}},\ \bibinfo
  {pages} {561} (\bibinfo {year} {2002})}\BibitemShut {NoStop}%
\bibitem [{\citenamefont {Fabrizio}\ \emph {et~al.}(1998)\citenamefont
  {Fabrizio}, \citenamefont {Altarelli},\ and\ \citenamefont
  {Benfatto}}]{Fabrizio1998}%
  \BibitemOpen
  \bibfield  {author} {\bibinfo {author} {\bibfnamefont {M.}~\bibnamefont
  {Fabrizio}}, \bibinfo {author} {\bibfnamefont {M.}~\bibnamefont
  {Altarelli}},\ and\ \bibinfo {author} {\bibfnamefont {M.}~\bibnamefont
  {Benfatto}},\ }\bibfield  {title} {\bibinfo {title} {{X-Ray Resonant
  Scattering as a Direct Probe of Orbital Ordering in Transition-Metal
  Oxides}},\ }\href {https://doi.org/10.1103/PhysRevLett.80.3400} {\bibfield
  {journal} {\bibinfo  {journal} {Phys. Rev. Lett.}\ }\textbf {\bibinfo
  {volume} {80}},\ \bibinfo {pages} {3400} (\bibinfo {year}
  {1998})}\BibitemShut {NoStop}%
\bibitem [{\citenamefont {Ament}\ \emph {et~al.}(2011)\citenamefont {Ament},
  \citenamefont {van Veenendaal}, \citenamefont {Devereaux}, \citenamefont
  {Hill},\ and\ \citenamefont {van~den Brink}}]{Ament2011}%
  \BibitemOpen
  \bibfield  {author} {\bibinfo {author} {\bibfnamefont {L.~J.~P.}\
  \bibnamefont {Ament}}, \bibinfo {author} {\bibfnamefont {M.}~\bibnamefont
  {van Veenendaal}}, \bibinfo {author} {\bibfnamefont {T.~P.}\ \bibnamefont
  {Devereaux}}, \bibinfo {author} {\bibfnamefont {J.~P.}\ \bibnamefont
  {Hill}},\ and\ \bibinfo {author} {\bibfnamefont {J.}~\bibnamefont {van~den
  Brink}},\ }\bibfield  {title} {\bibinfo {title} {Resonant inelastic x-ray
  scattering studies of elementary excitations},\ }\href
  {https://doi.org/10.1103/RevModPhys.83.705} {\bibfield  {journal} {\bibinfo
  {journal} {Rev. Mod. Phys.}\ }\textbf {\bibinfo {volume} {83}},\ \bibinfo
  {pages} {705} (\bibinfo {year} {2011})}\BibitemShut {NoStop}%
\end{thebibliography}

%

\end{document}